\documentclass[onecolumn]{aa}

\usepackage{graphicx}

\begin{document}
\title{On Quasilinear Perpendicular Diffusion}

\author{O. Stawicki}

\institute{Unit for Space Physics, School of
	Physics, North-West University, Potchefstroom 2520, South Africa}

\date{Received  / Accepted }

\authorrunning{O. Stawicki}
\titlerunning{On quasilinear perpendicular diffusion}

\abstract{
Quasilinear perpendicular diffusion of charged particles in
fluctuating electromagnetic fields is the focus of this paper. A
general transport parameter for perpendicular diffusion is presented being valid for an arbitrary
turbulence geometry and a plasma wave dispersion relation varying
arbitrarily in wavevector. 
The new diffusion coefficient is evaluated in detail for slab
turbulence geometry for two special cases: 
(1) Alfv\'enic turbulence and (2) dynamical magnetic 
turbulence. Furthermore, perpendicular diffusion in 2D geometry is
considered for a purely dynamical magnetic turbulence.
The derivations and numerical calculations presented here cast serious
doubts on the applicability of quasilinear theory for perpendicular
diffusion. Furthermore, they emphasize that nonlinear effects play a
crucial role in the context of perpendicular diffusion.
}

\maketitle

\section{Introduction and Motivation}
The influence of a turbulence on the spatial transport of charged
particles plays a key role in a variety of heliospheric and
astrophysical scenarios. The knowledge of the properties of the 
underlying turbulence is crucial for understanding the 
three-dimensional (anisotropic) diffusive particle transport in a
collisionless, turbulent and magnetized plasma such as the solar wind
or the interstellar medium. Of particular interest are the transport
coefficients describing particle diffusion perpendicular to an ambient 
magnetic field. 

In spite of its long-standing importance in space and astrophysics,
perpendicular diffusion has been an unsolved puzzle during many
decades and a variety of studies have been carried out to achieve closure 
and to pin it down at a theoretical level. Models have been proposed
using hard-sphere scattering in a magnetized plasma (Gleeson
\cite{gle}) and extended version of it developed on the basis of the 
Boltzmann equation
(Jones \cite{jones1990}), but they are inapplicable to space plasmas
where (electro)magnetic fluctuations trigger particle
scattering. Other models are based on the field line random walk
(FLRW) limit emerging, for slab turbulence geometry, from quasilinear theory
(QLT; Jokipii \cite{jo1}; Jokipii \& Parker \cite{jo2}), which has
been considered and used in subsequent studies 
(e.g., Forman et al. \cite{for1}; Forman \cite{for2}; Bieber \&
Matthaeus \cite{biemat97}). Although FLRW provides for a
physically appealing picture, it has been shown recently that its
applicability is questionable for particle transport in certain
turbulence geometries, particularly those with at least one ignorable
coordinate being the case for slab geometry (Jokipii et al. \cite{jok93}). 
Furthermore, numerical simulations, taking into account the magnetic
nature of a turbulence, have shown that FLRW fails to explain perpendicular
transport of low-energy particles  (Giacalone \& Jokipii
\cite{giajok99}; Mace et al. \cite{mace}), nor is it clear
that it actually accounts for the cross-field transport of charged particles.  

While perpendicular and parallel particle transport processes are necessarily
distinct from one another in QLT, it has been argued for some time that 
parallel particle scattering can reduce perpendicular diffusion to
subdiffusive levels if the turbulence reveals slab geometry only
(see, e.g., Urch \cite{urch77}; K\'ota \& Jokipii
\cite{kotajokipii2000}; Qin et al. \cite{qinetal2002grl}). However, when the
turbulent magnetic field has sufficient structure normal to the mean
magnetic field, subdiffusion as seen in pure slab turbulence can be
overcome and diffusion is recovered (see Qin et al. \cite{qin}). 

Recently, Matthaeus et al. (\cite{mat03}) proposed a promising model
for perpendicular diffusion, also referred to as nonlinear guiding
center (NLGC) model. The model takes into account different
turbulence geometries and parallel diffusion while the particle
gyrocenter follows magnetic field lines. Based on their
model and its comparison with numerical simulations, it became clear that
nonlinear effects play a crucial role for a more realistic
understanding of perpendicular diffusion.

Although more advanced models such as the NLGC approach have been
developed, it is nevertheless instructive to provide a rigorous
treatment of perpendicular diffusion in QLT. Related calculations
by Shalchi \& Schlickeiser (\cite{shalchischlickeiser2004}) are valid
for purely magnetic fields only. Furthermore, their calculations are
restricted to specific  wavenumber variations of the magnetic power
spectrum and the turbulence correlation time. Detailed calculations
do not exist allowing to consider quasilinear perpendicular diffusion in
an electromagnetic turbulence.

In view of this lack, it is desirable to derive a
general QLT Fokker-Planck coefficient for an electromagnetic
turbulence with arbitrary geometry. This and the detailed
evaluation of the new coefficient for the limit of slab and 2D
geometry is the purpose of this paper.

The structure of this paper is as follows: Section \ref{sec:deriv}
introduces the governing quasilinear equations of motions for test
particles in fluctuating electromagnetic fields, and a general
Fokker-Planck coefficient for a plasma wave turbulence with arbitrary
geometry and arbitrary wave dispersion relation is derived. The
limit of a slab turbulence geometry is considered in detail in section
\ref{sec:slab}. There, perpendicular diffusion coefficients are
presented for (1) an
Alfv\'enic turbulence and (2) a purely magnetic turbulence. The limit
of 2D turbulence geometry is considered in section
\ref{sec:2d}. Numerical calculations and conclusions for both the slab
and 2D geometry are presented in section \ref{sec:num}.

\section{Quasilinear Derivation of the General Diffusion Coefficient}
\label{sec:deriv}
An often used standard approach for the evaluation of spatial
diffusion coefficients is to calculate them from an ensemble of
particle trajectories. To do so, the Taylor-Green-Kubo (TGK) formula
is often applied. For the diffusion coefficient in $x$-direction, the
TGL formula (Kubo \cite{kubo1957}) reads
\begin{equation}
\kappa_{xx}=\int\limits_{0}^{t}d\xi<v_x(0)v_x(\xi)>
\label{eq:tgk1}
\end{equation}
in the limit $t\to\infty$, where $v_x$ is the $x$-component of the
particle velocity. The brackets $<\ldots>$ denote an ensemble average
over the relevant two-time distribution of particles. An analogous
expression holds for the diffusion coefficient in $y$-direction. For a large
coherence time $\xi$, the second-order velocity correlation function
$<v_x(0)v_x(\xi)>$ must go to zero, and the integral in
Eq.(\ref{eq:tgk1}) approaches a constant value for
$t\to\infty$. Following the argumentation done by Kubo (see also
K\'ota \& Jokipii \cite{kotajokipii2000}), equation (\ref{eq:tgk1}) results from
\begin{equation}
<\Delta x^2>=\left<\left(\int\limits_{0}^{t}d\xi v_x(\xi)\right)^2\right>=2\int\limits_{0}^{t}d\xi(t-\xi)<v_x(0)v_{x}(\xi)>
\label{eq:tgk2}
\end{equation}
where $<\Delta x^2>$ is the mean square displacement of the particle
position in time $t$. For $t$ large compared to $\xi$,
Eq. (\ref{eq:tgk2}) leads to $<\Delta x^2>=2\kappa_{xx}t$, with the
diffusion coefficient $\kappa_{xx}$ given by equation (\ref{eq:tgk1}).

The use of the TGK formula is somewhat critical for situations
where particle transport reveals rather an anomalous diffusion
process than normal Markovian diffusion. In this case, the standard
Fokker-Planck coefficient (anomalous transport law) 
\begin{equation}
\kappa_{xx}=\frac{<\Delta x^2>}{2t}
\label{eq:adc}
\end{equation}
applies, where the mean square displacement scales more general as
$<\Delta x^2>\propto t^{\gamma}$. Depending on the exponent $\gamma$,
different diffusion processes can be taken into account: $\gamma=1$ for
the diffusive regime, corresponding to a Gaussian random walk of particles;
$\gamma=2$ in the superdiffusive regime, i.e. strictly scatter-free
propagation of the particles, and $\gamma<1$ (e.g. $\gamma=1/2$, as
shown by Qin et al. \cite{qinetal2002grl} and Kota \& Jokipii \cite{kotajokipii2000}) in the case of particle
trapping (subdiffusion or compound diffusion). For $\gamma=1$ (normal
Markovian diffusion), the anomalous transport law (\ref{eq:adc}) is
equivalent to the TGK formula (\ref{eq:tgk1}) for the large $t$
limit (see also Bieber \& Matthaeus \cite{biemat97}).

In the context of QLT, the quasilinear perpendicular diffusion
coefficients $\kappa_{XX}$ and $\kappa_{YY}$ can be written as
(Schlickeiser \cite{sch02}, Eqs.[12.3.25] and [12.3.26])
\begin{equation}
\kappa_{XX}=\frac{1}{2}\int\limits_{-1}^{1}d\mu D_{XX}
\,\,\,;\hspace*{0.3cm}
\kappa_{YY}=\frac{1}{2}\int\limits_{-1}^{1}d\mu D_{YY}
\label{eq:tgk}
\end{equation}
where $\mu=v_{\|}/v$ is the pitch-angle of a particle having the
velocity component $v_{\|}$ along the ordered magnetic field
$B_0$. The subscripts $X$ and $Y$ denote guiding center coordinates,
and $D_{XX}$ and $D_{YY}$ are Fokker-Planck coefficients
(cf. Schlickeiser \cite{sch02}, Eq. [12.1.29]) of the form
\begin{equation}
D_{XX}=\Re\int\limits_{0}^{\infty}d\xi<{\dot{X}}(t){\dot{X}}^{\ast}(t+\xi)>
\,\,\,;\hspace*{0.3cm}
D_{YY}=\Re\int\limits_{0}^{\infty}d\xi<{\dot{Y}}(t){\dot{Y}}^{\ast}(t+\xi)>
\label{eq:fpcdef}
\end{equation}
They represent the interaction of a particle with electromagnetic
fluctuations. The relations as given in equation (\ref{eq:fpcdef})
correspond to the TGK formula (\ref{eq:tgk1}) used earlier by Bieber \&
Matthaeus (\cite{biemat97}) (see their Eq. [2]). In QLT, however, one
has to perform an additional average with respect to
$\mu$. If the fluctuations are statistically homogeneous in space and
time $t$, the velocity correlation functions in equation
(\ref{eq:fpcdef}) depend only on $\xi$. This can simply be taken into
account in the Fokker-Planck coefficients (\ref{eq:fpcdef}) by setting
$t=0$, and the quasilinear relations (\ref{eq:fpcdef}) then
correspond, apart from the $\mu$-integration, to the TGK formula
(\ref{eq:tgk1}).

On the basis of Eq. (\ref{eq:fpcdef}), QLT does not allow to consider
subdiffusion or superdiffusion, since the derivations of
the relations (\ref{eq:fpcdef}) are based on the assumption, that $t$
is larger than the coherence time $\xi$ (see Schlickeiser
\cite{sch02}, Eq. [12.1.17] and the comments following it). However,
it was shown recently by K\'ota \& Jokipii (\cite{kotajokipii2000})
 that the Kubo
formalism does not necessarily contradict compound (sub)diffusion if
modifications are applied. The consideration of
anomalous particle diffusion ($\gamma\neq 1$), particularly of
subdiffusion and an associated modification of the Kubo formalism, is
beyond the purview of this paper and normal diffusion is simply
assumed by applying the TGK approach.

\subsection{Equations of Motion}
\label{sec:eom}
The determination of $D_{XX}$ and $D_{YY}$ requires the knowledge of
the equations of motion. According to Schlickeiser (\cite{sch02}) (see
his Eqs. [12.1.9d] and [12.1.9e]), the perpendicular components of the
fluctuating force fields can be written as
\begin{eqnarray}
g_X = \dot X(t) & = & -v\cos\phi(t)\sqrt{1-\mu^2}{\delta B_{\|}\over B_0}
+{c\over B_0}\left(\delta E_y+\mu{v\over c}\delta B_x\right)
\label{eq:fffx}
\\[0.2cm]
g_Y =\dot Y(t) & = & -v\sin\phi(t)\sqrt{1-\mu^2}{\delta B_{\|}\over
  B_0}-{c\over B_0}\left(\delta E_x-\mu{v\over c}\delta B_y\right)
\label{eq:fffy}
\end{eqnarray}
where $\phi$ denotes the gyrophase of the particle. Note that the
Cartesian components of the fluctuating electromagnetic field,
i.e. $\delta B_{x,y,\|}$ and $\delta E_{x,y,\|}$, are used and
not the helical description, i.e. left- and right-hand polarized
fields.

For the further treatment of equations (\ref{eq:fffx}) and
(\ref{eq:fffy}), a standard perturbation method is applied. To do so,
it is convenient to replace in the Fourier transform of the irregular
electromagnetic field the true particle orbit ${\bf x}(t)$ by an
unperturbed particle orbit, yielding
\begin{equation}
{\bf \delta B}({\bf x},t)=\int\limits_{}^{}d^3k{\bf\delta B}({\bf k},t)e^{\imath {\bf x}(t)\cdot t}=
\sum\limits_{n=-\infty}^{+\infty}\int\limits_{}^{}d^3k {\bf \delta
  B}({\bf k},t)J_{n}(W)\exp\biggl[\imath n[\psi-\phi(t)]+\imath
  k_{\|}v_{\|}t\biggr]
\label{eq:fourier}
\end{equation}
and an analogous expression for $\delta {\bf E}$. The quantity
$J_n(W)$ is a Bessel function of the first kind and order $n$. The
particle gyrophase for an unperturbed orbit is given by
$\mbox{$\phi(t)=\phi_0-\Omega t$}$, where the random variable $\phi_0$
denotes the initial gyrophase of the particle. Furthermore, the
abbreviation $W=k_{\perp}R_L\sqrt{1-\mu^2}$ is introduced, where
$R_L=v/\Omega$ is the Larmor radius. The relativistic gyrofrequency is
given by $\Omega=qB_0/(\gamma mc)$ with $m$ being the mass and $q$ the
charge of the particle, $\gamma$ is the Lorentz factor. The angle
$\psi$ results from the wavenumber representation
$k_x=k_{\perp}\cos\psi$ and $k_y=k_{\perp}\sin\psi$. With equation
(\ref{eq:fourier}), the equations of motion (\ref{eq:fffx}) and
(\ref{eq:fffy}) can be manipulated to become
\begin{eqnarray}
{\dot{X}}(t)& = & {v\over B_0}\sum_{n=-\infty}^{\infty}\int\limits_{}^{}d^3 k \exp\biggl[\imath n\left[\psi-\phi(t)\right]+\imath k_{\|}v_{\|}t\biggr]
\label{eq:fffx1}
\\[0.2cm]
&&
\times\left\{-\frac{\sqrt{1-\mu^2}}{2}\Biggl[J_{n+1}(W)e^{\imath\psi}+e^{-\imath\psi}J_{n-1}(W)\Biggr]\delta B_{\|}
+{c\over v}J_n(W)\left(\delta E_y+\mu{v\over c}\delta B_x\right)\right\}
\nonumber
\\[0.4cm]
{\dot{Y}}(t) & = & {v\over B_0}\sum_{n=-\infty}^{\infty}\int\limits_{}^{}d^3 k \exp\biggl[\imath n\left[\psi-\phi(t)\right]+\imath k_{\|}v_{\|}t\biggr]
\label{eq:fffy1}
\\[0.2cm]
&&
\times\left\{\imath\frac{\sqrt{1-\mu^2}}{2}\Biggl[J_{n+1}(W)e^{\imath\psi}-e^{-\imath\psi}J_{n-1}(W)\Biggr]\delta B_{\|}-{c\over v}J_n(W)\left(\delta E_x-\mu{v\over c}\delta B_y\right)\right\}
\nonumber
\end{eqnarray}
For the evaluation of equation (\ref{eq:tgk}), it is convenient to
consider now the nature of the electromagnetic turbulence. Here, the 
``wave viewpoint'' is used by assuming that the
turbulence can be represented by a superposition of $N$ individual plasma
wave modes, so that
\begin{equation}
\delta {\bf B}({\bf k},t)=\sum\limits_{j=1}^{N}\delta {\bf B}^j({\bf k})\exp(-\imath \omega_j t)\,\,;
\hspace*{0.2cm}
\delta {\bf E}({\bf k},t)=\sum\limits_{j=1}^{N}\delta {\bf E}^j({\bf k})\exp(-\imath \omega_j t)
\label{eq:super}
\end{equation}
Here, $\omega_j({\bf k})=\omega_{j,R}({\bf k})+\imath\Gamma_j({\bf
  k})$ is a complex dispersion relation of wave
mode $j$, where $\omega_{j,R}$ is the real frequency of the mode. The
  imaginary part, $\Gamma_j({\bf k})\leq 0$, represents dissipation of
  turbulent energy due to plasma wave damping.

Restricting the considerations to transverse fluctuations, i.e. $\delta {\bf E}^j\cdot {\bf k}=0$, and using Faraday's law, the turbulent electric field can easily be expressed by the corresponding magnetic counterparts, yielding
\begin{equation}
\delta E_x^j=\frac{\omega_j}{ck^2}\left(\delta B_y^jk_{\|}-\delta B_{\|}^jk_y\right)
\,\,\,;\hspace*{0.1cm}
\delta E_y^j=\frac{\omega_j}{ck^2}\left(\delta B_{\|}^jk_x-\delta B_x^jk_{\|}\right)
\label{eq:fara}
\end{equation}
Having expressed the electric by the magnetic field, it is also convenient to introduce now the Bessel function identities 
\begin{equation}
J_{n-1}(W)+J_{n+1}(W)= \frac{2n}{W}J_{n}(W)
\,\,\,;\hspace*{0.3cm}
J_{n-1}(W)-J_{n+1}(W) =2J_{n}^{\prime}(W)
\label{eq:besselid}
\end{equation}
where the prime denotes the derivation with respect to $W$. With equations (\ref{eq:fara}) and (\ref{eq:besselid}), the equations of motion (\ref{eq:fffx1}) and (\ref{eq:fffy1}) can readily be rearranged, and one arrives at
\begin{eqnarray}
{\dot{X}}(t)& = & -{v\over B_0}\sum\limits_{j}^{}\sum_{n=-\infty}^{\infty}\int\limits_{}^{}d^3 k \exp\biggl[\imath n\left[\psi-\phi(t)\right]+\imath (k_{\|}v_{\|}-\omega_j)t\biggr]
\label{eq:fffx2}
\\[0.2cm]
&&
\times\left\{J_{n}(W)\left[a\frac{k_x}{k_{\perp}}\delta B_{\|}^j+b\delta B_x^j\right]-\imath\sqrt{1-\mu^2}\frac{k_y}{k_{\perp}}\delta B_{\|}^jJ_{n}^{\prime}(W)\right\}
\nonumber
\\[0.4cm]
{\dot{Y}}(t) & = & -{v\over B_0}\sum\limits_{j=1}^{N}\sum_{n=-\infty}^{\infty}\int\limits_{}^{}d^3 k \exp\biggl[\imath n\left[\psi-\phi(t)\right]+\imath (k_{\|}v_{\|}-\omega_j)t\biggr]
\label{eq:fffy2}
\\[0.2cm]
&&
\times\left\{J_{n}(W)\left[a\frac{k_y}{k_{\perp}}\delta B_{\|}^j+b\delta B_y^j\right]+\imath\sqrt{1-\mu^2}\frac{k_x}{k_{\perp}}\delta B_{\|}^jJ_{n}^{\prime}(W)\right\}
\nonumber
\end{eqnarray}
where the following complex functions have been introduced:
\begin{equation}
a=\frac{n}{W}\sqrt{1-\mu^2}-\frac{\omega_jk_{\perp}}{v\,k^2}
\,\,\,;\hspace*{0.3cm}
b=\frac{\omega_jk_{\|}}{v\,k^2}-\mu
\label{eq:perpabini}
\end{equation}

\subsection{Velocity Correlation Functions}
\label{sec:ccf}
Having determined the equations of motion, one can now proceed to
calculate the second-order velocity correlation functions
$<{\dot{X}}(t){\dot{X}}^{\ast}(t+\xi)>$ and
$<{\dot{Y}}(t){\dot{Y}}^{\ast}(t+\xi)>$ entering equation
(\ref{eq:fpcdef}). The procedure for the calculation is relatively
lengthy, but can be carried out with simple
algebra. For the sake of generality, the explicit time dependence of
the correlation functions is included. At the end of this section,
the limit $t=0$ is considered to obtain
$<{\dot{X}}(0){\dot{X}}^{\ast}(\xi)>$ and 
$<{\dot{Y}}(0){\dot{Y}}^{\ast}(\xi)>$, required for statistically
homogeneous conditions.

The calculations for both velocity correlation functions are
analogous, and the calculations are, therefore, restricted to
$\mbox{$<{\dot{X}}(t){\dot{X}}^{\ast}(t+\xi)>$}$. Multiplying equation
(\ref{eq:fffx2}) with its conjugated leads to
\begin{eqnarray}
{\dot{X}}(t){\dot{X}}^{\ast}(t+\xi) & = & \frac{v^2}{B_0^2}\sum_{j}^{}\sum_{n=-\infty}^{\infty}\sum_{m=-\infty}^{\infty}\int\limits_{}^{}d^3 k\int\limits_{}^{}d^3{\overline k}\,\exp(\chi)
\label{eq:corrfun}
\\
&&\times\Biggl\{J_{n}(W)J_{m}(\overline{W})\biggl[a\overline a^{\ast}\frac{k_x\overline k_x}{k_{\perp}\overline k_{\perp}}\cdot (\delta B_{\|}^j\delta {\overline{B}}_{\|}^{j\ast})+b\overline b^{\ast}\cdot (\delta B_{x}^j\delta \overline B_{x}^{j\ast})
\nonumber
\\
&&\hspace*{3.5cm}+b\overline a^{\ast}\frac{\overline k_x}{\overline k_{\perp}}\cdot (\delta B_{x}^j\delta \overline B_{\|}^{j\ast})+a\overline b^{\ast}\frac{k_x}{k_{\perp}}\cdot (\delta B_{\|}^j\delta \overline B_{x}^{j\ast})\biggr]
\nonumber
\\
&&
+\imath\sqrt{1-\mu^2}J_{n}(W)J_{m}^{\prime}(\overline W)\frac{\overline k_y}{\overline k_{\perp}}\left[a\frac{k_x}{k_{\perp}}\cdot(\delta B_{\|}^j\delta \overline B_{\|}^{j\ast})+b\cdot(\delta B_{x}^j\delta\overline B_{\|}^{j\ast})\right]
\nonumber
\\
&&
-\imath\sqrt{1-\mu^2}J_{m}(\overline{W})J_{n}^{\prime}(W)\frac{k_y}{k_{\perp}}\left[\overline a^{\ast}\frac{\overline k_x}{\overline k_{\perp}}\cdot(\delta B_{\|}^j\delta \overline B_{\|}^{j\ast})+\overline b^{\ast} \cdot(\delta B_{\|}^j\delta \overline B_{x}^{j\ast})\right]
\nonumber
\\
&&+(1-\mu^2)\frac{k_y{\overline k_y}}{k_{\perp}{\overline k_{\perp}}}J_{n}^{\prime}(W)J_{m}^{\prime}(\overline{W})\cdot(\delta B_{\|}^j\delta \overline B_{\|}^{j\ast})\Biggr\}
\nonumber
\end{eqnarray}
where the abbreviation
\begin{equation}
\chi=\imath (n\psi-m\overline{\psi})-\imath(n-m)\phi(t)+\imath v_{\|}(k_{\|}-{\overline k_{\|}})t+\imath(\overline\omega_j^{\ast}-\omega_j)t-\imath({\overline k_{\|}}v_{\|}-\overline\omega_j^{\ast}+m\Omega)\xi
\end{equation}
has been introduced. The bar notation used over some quantities
indicates that they have to be evaluated for wavevector $\overline
{\bf k}$ and time $t+\xi$. A simplification of (\ref{eq:corrfun}) can
be achieved if one applies an average with respect to the random
variable $\phi_0$, the initial gyrophase of the particle. Using the
relation
\begin{equation}
\frac{1}{2\pi}\int\limits_{0}^{2\pi}d\phi_0\exp[\imath(n-m)\phi_0]=\delta_{nm}
\end{equation}
where Kronecker's $\delta_{nm}=0$ for $n\neq m$ and unity for $n=m$, and applying the ensemble average, equation (\ref{eq:corrfun}) can be manipulated to become
\begin{eqnarray}
<{\dot{X}}(t){\dot{X}}^{\ast}(t+\xi)> & = & \frac{v^2}{B_0^2}\sum_{j}^{}\sum_{n=-\infty}^{\infty}\sum_{m=-\infty}^{\infty}\int\limits_{}^{}d^3 k\int\limits_{}^{}d^3{\overline k}\,\exp(\chi)
\label{eq:corrfun2}
\\
&&\times\Biggl\{J_{n}(W)J_{n}(\overline{W})\biggl[a\overline a^{\ast}\frac{k_x\overline k_x}{k_{\perp}\overline k_{\perp}}<\delta B_{\|}^j\delta \overline B_{\|}^{j\ast}>+b\overline b^{\ast}<\delta B_{x}^j\delta \overline B_{x}^{j\ast}>
\nonumber
\\
&&\hspace*{3.5cm}+b\overline a^{\ast}\frac{\overline k_x}{\overline k_{\perp}}<\delta B_{x}^j\delta \overline B_{\|}^{j\ast}>+a\overline b^{\ast}\frac{k_x}{k_{\perp}}<\delta B_{\|}^j\delta \overline B_{x}^{j\ast}>\biggr]
\nonumber
\\
&&
+\imath\sqrt{1-\mu^2}J_{n}(W)J_{n}^{\prime}(\overline W)\frac{\overline k_y}{\overline k_{\perp}}\left[a\frac{k_x}{k_{\perp}}<\delta B_{\|}^j\delta \overline B_{\|}^{j\ast}>+b<\delta B_{x}^j\delta\overline B_{\|}^{j\ast}>\right]
\nonumber
\\
&&
-\imath\sqrt{1-\mu^2}J_{n}(\overline{W})J_{n}^{\prime}(W)\frac{k_y}{k_{\perp}}\left[\overline a^{\ast}\frac{\overline k_x}{\overline k_{\perp}}<\delta B_{\|}^j\delta \overline B_{\|}^{j\ast}>+\overline b^{\ast} <\delta B_{\|}^j\delta \overline B_{x}^{j\ast}>\right]
\nonumber
\\
&&+(1-\mu^2)\frac{k_y{\overline k_y}}{k_{\perp}{\overline k_{\perp}}}J_{n}^{\prime}(W)J_{n}^{\prime}(\overline{W})<\delta B_{\|}^j\delta \overline B_{\|}^{j\ast}>\Biggr\}
\nonumber
\end{eqnarray}
where the quantity $\chi$ is now given by 
\begin{equation}
\chi=\imath n(\psi-\overline{\psi})+\imath v_{\|}(k_{\|}-{\overline k_{\|}})t+\imath(\overline\omega_j^{\ast}-\omega_j)t-\imath({\overline k_{\|}}v_{\|}-\overline\omega_j^{\ast}+n\Omega)\xi
\end{equation}
The next step consists of the assumption that Fourier components at
different wave vectors are uncorrelated. Introducing the subscripts
$\alpha$ and $\beta$ for Cartesian coordinates, the ensemble averages
of the magnetic field fluctuations can also be written as
\begin{equation}
<\delta B_{\alpha}^j\delta \overline {B}_{\beta}^{j\ast}>=<\delta B_{\alpha}^j({\bf k})\delta B_{\beta}^{j\ast}({\bf k}^{\prime})>=\delta({\bf k}-{\bf k}^{\prime})P_{\alpha\beta}^j({\bf k})
\label{eq:mcorin}
\end{equation}
The uncorrelated state implies $\psi=\overline\psi$, $W=\overline W$
and $\omega_j=\overline\omega_j$, and the velocity correlation
function, equation (\ref{eq:corrfun2}), reduces to 
\begin{eqnarray}
<{\dot{X}}(t){\dot{X}}^{\ast}(t+\xi)> & = & \frac{v^2}{B_0^2}\sum_{j}^{}\sum_{n=-\infty}^{\infty}\int\limits_{}^{}d^3 k\,\exp\left[2\Gamma_jt-\imath(k_{\|}v_{\|}-\omega_{j,R}+n\Omega)\xi+\Gamma_j\xi\right]
\label{eq:corrfunxx}
\\
&&\times\Biggl\{J_{n}^2(W)\biggl[aa^{\ast}\frac{k_x^2}{k_{\perp}^2}P_{\|\|}^j+bb^{\ast}P_{xx}^j+\frac{k_x}{k_{\perp}}\left[ab^{\ast}P_{\| x}^j+ba^{\ast} P_{x\|}^j\right]\biggr]
\nonumber
\\
&&+\imath\sqrt{1-\mu^2}J_{n}(W)J_{n}^{\prime}(W)\frac{k_y}{k_{\perp}}\biggl[(a-a^{\ast})\frac{k_x}{k_{\perp}}P_{\|\|}^j+bP_{x\|}^j-b^{\ast}P_{\|x}^j\biggr]
\nonumber
\\
&&+(1-\mu^2)\frac{k_y^2}{k_{\perp}^2}\left[J_{n}^{\prime}(W)\right]^2P_{\|\|}^j\Biggr\}
\nonumber
\end{eqnarray}
Analogous to the calculations presented above, one obtains for the velocity correlation function in $Y$-direction the expression
\begin{eqnarray}
<{\dot{Y}}(t){\dot{Y}}^{\ast}(t+\xi)> & = & \frac{v^2}{B_0^2}\sum_{j}^{}\sum_{n=-\infty}^{\infty}\int\limits_{}^{}d^3 k\,\exp\left[2\Gamma_jt-\imath(k_{\|}v_{\|}-\omega_{j,R}+n\Omega)\xi+\Gamma_j\xi\right]
\label{eq:corrfunyy}
\\
&&\times\Biggl\{J_{n}^2(W)\biggl[aa^{\ast}\frac{k_y^2}{k_{\perp}^2}P_{\|\|}^j+bb^{\ast}P_{yy}^j+\frac{k_y}{k_{\perp}}\left[ab^{\ast}P_{\| y}^j+ba^{\ast} P_{y\|}^j\right]\biggr]
\nonumber
\\
&&-\imath\sqrt{1-\mu^2}J_{n}(W)J_{n}^{\prime}(W)\frac{k_x}{k_{\perp}}\biggl[(a-a^{\ast})\frac{k_y}{k_{\perp}}P_{\|\|}^j+bP_{y\|}^j-b^{\ast}P_{\|y}^j\biggr]
\nonumber
\\
&&+(1-\mu^2)\frac{k_x^2}{k_{\perp}^2}\left[J_{n}^{\prime}(W)\right]^2P_{\|\|}^j\Biggr\}
\nonumber
\end{eqnarray}
The advantage of having applied the Bessel function identities
(\ref{eq:besselid}) to the equations of motion (\ref{eq:fffx2}) and
(\ref{eq:fffy2}) is now obvious. Both velocity correlation functions
and, therefore, $\kappa_{XX}$ and $\kappa_{YY}$ are expressed as a sum
of three terms. Each contribution includes either $J_{n}^2(W)$,
$J_{n}(W)J_{n}^{\prime}(W)$ or $[J_{n}^{\prime}(W)]^2$. Each term
is accompanied by a specific factor which contains the components of
the magnetic correlation tensor $P_{\alpha\beta}^j({\bf k})$ and,
therefore, the complete information about the turbulence geometry. 

Furthermore, note that the velocity correlation functions reveal an
explicit dependence on time $t$. This is a consequence of the
dissipation of turbulent energy due to plasma wave damping. To
demonstrate this, equation (\ref{eq:super}) is considered again but,
for simplicity, without the concept of a superposition of different
wave modes, so that $\delta {\bf B}({\bf k},t)=\delta {\bf B}({\bf k})\exp(-\imath \omega t)$
where $\omega=\omega_R+\imath\Gamma$ with the dissipation rate $\Gamma\leq 0$.
The corresponding turbulent energy can then be expressed as
\begin{equation}
<\delta B({\bf k},t)\delta B^{\ast}({\bf k},t+\xi)>=P({\bf k},t,\xi)=<\delta B({\bf k})\delta B^{\ast}({\bf
  k})>e^{(\imath\omega_R+\Gamma)\xi}e^{2\Gamma t}=P({\bf k},\xi)e^{2\Gamma t}
\end{equation}
yielding, for a dissipative system, the well known relation
\begin{equation}
\frac{dP({\bf k},t,\xi)}{dt}=2\Gamma P({\bf k},t,\xi)
\end{equation}
where $2\Gamma$ is the rate of energy dissipation due to collisionless
plasma wave damping. According to the TGK formula
(\ref{eq:tgk1}), one has to evaluate the velocity correlation
functions for the condition $t=0$,
e.g. $<{\dot{X}}(0){\dot{X}}^{\ast}(\xi)>$, since the fluctuations 
are assumed to be homogeneous in space and time, and the variation in
$t$ vanishes. Hence, the calculations presented below do not take
into account the contribution $2\Gamma_j t$. However, such a
contribution may be of interest for a more general treatment, whether
in QLT or in an extendet version of the NLGC model, which is not
considered here.

\subsection{Fokker-Planck Coefficients}
Having determined the velocity correlation functions in the previous
section, one can now proceed and evaluate the Fokker-Planck
coefficients as given by the relations in equation (\ref{eq:fpcdef})
with $t=0$ in the velocity correlation functions. Upon
substituting the expressions (\ref{eq:corrfunxx}) and
(\ref{eq:corrfunyy}), one obtains
\begin{equation}
D_{XX}=\frac{v^2}{B_0^2}
\sum\limits_{j}^{}
\sum\limits_{n=-\infty}^{\infty}\Re
\int\limits_{}^{}d^3 k {\cal R}_{j}
\Biggl[J_{n}^2(W)F_{X}^j+\imath J_{n}(W)J_{n}^{\prime}(W)G_{X}^j+\bigl[J_{n}^{\prime}(W)\bigr]^2H_{X}^j\Biggr]
\label{eq:dxx}
\end{equation}
and an analogous expression for $D_{YY}$, but including different
auxiliary functions $F_{Y}^j$, $G_{Y}^j$ and $H_{Y}^j$. The latter and
their corresponding counterparts for $D_{XX}$ read as follows:
\begin{eqnarray}
F_{X}^j
& = &  aa^{\ast}\frac{k_x^2}{k_{\perp}^2}P_{\|\|}^j+bb^{\ast}P_{xx}^j+\frac{k_x}{k_{\perp}}\left[ab^{\ast}P_{\| x}^j+ba^{\ast} P_{x\|}^j\right]
\label{eq:auxbeg}
\\[0.5cm]
F_{Y}^j
& = &  
aa^{\ast}\frac{k_y^2}{k_{\perp}^2}P_{\|\|}^j+bb^{\ast}P_{yy}^j+\frac{k_y}{k_{\perp}}\left[ab^{\ast}P_{\| y}^j+ba^{\ast} P_{y\|}^j\right]
\\[0.5cm]
G_{X}^j
& = & 
\sqrt{1-\mu^2}\frac{k_y}{k_{\perp}}\biggl[(a-a^{\ast})\frac{k_x}{k_{\perp}}P_{\|\|}^j+bP_{x\|}^j-b^{\ast}P_{\|x}^j\biggr]
\\[0.5cm]
G_{Y}^j
& = &
-\sqrt{1-\mu^2}\frac{k_x}{k_{\perp}}\biggl[(a-a^{\ast})\frac{k_y}{k_{\perp}}P_{\|\|}^j+bP_{y\|}^j-b^{\ast}P_{\|y}^j\biggr]
\\[0.5cm]
H_{X}^j
& = & \left(1-\mu^2\right)\frac{k_y^2}{k_{\perp}^2}P_{\|\|}^j
\\[0.5cm]
H_{Y}^j
& = & \left(1-\mu^2\right)\frac{k_x^2}{k_{\perp}^2}P_{\|\|}^j
\label{eq:auxend}
\end{eqnarray}
The integration with respect to $\xi$ leads to the complex resonance function,
\begin{eqnarray}
{\cal R}_{j} & = & \int\limits_{0}^{\infty}d\xi
\exp\left[-\imath(k_{\|}v_{\|}-\omega_{j,R}+n\Omega)\xi+\Gamma_j\xi\right]
\nonumber
\\
& = & 
-\frac{\Gamma_j+\imath(k_{\|}v_{\|}-\omega_{j,R}+n\Omega)}{\Gamma_j^2+(k_{\|}v_{\|}-\omega_{j,R}+n\Omega)^2}
\label{eq:respwg}
\end{eqnarray}
which describes interactions of the particles with the plasma wave
turbulence (remember that $\Gamma_j\leq 0$). 

The coefficients $D_{XX}$ and $D_{YY}$, as given by equation
(\ref{eq:dxx}), can also be summarized to yield a net coefficient,
i.e. $D_{\perp}=D_{XX}+D_{YY}$. According to equation (\ref{eq:tgk}),
one can then define a net perpendicular diffusion coefficient
\begin{equation}
\kappa_{\perp}=\kappa_{XX}+\kappa_{YY}=\frac{1}{2}\int\limits_{-1}^{+1}d\mu D_{\perp}
\label{eq:kappaperp}
\end{equation}
with
\begin{equation}
D_{\perp}=\frac{v^2}{B_0^2}
\sum\limits_{j}^{}
\sum\limits_{n=-\infty}^{\infty}\Re
\int\limits_{}^{}d^3 k {\cal R}_{j}
\Biggl[J_{n}^2(W)F_{\perp}^j+\imath J_{n}(W)J_{n}^{\prime}(W)G_{\perp}^j+\bigl[J_{n}^{\prime}(W)\bigr]^2H_{\perp}^j\Biggr]
\label{eq:dperp}
\end{equation}
The auxiliary functions $F_{\perp}^j$, $G_{\perp}^j$ and $H_{\perp}^j$
are then a simple superposition of the corresponding functions given
by equations (\ref{eq:auxbeg}) to (\ref{eq:auxend}), resulting in
\begin{eqnarray}
F_{\perp}^j\
& = & 
aa^{\ast}P_{\|\|}^j+bb^{\ast}\left(P_{xx}^j+P_{yy}^j\right)
+\frac{ab^{\ast}}{k_{\perp}}\left[k_xP_{\|x}^j+k_yP_{\|y}^j\right]
+\frac{ba^{\ast}}{k_{\perp}}\left[k_xP_{x\|}^j+k_yP_{y\|}^j\right]
\label{eq:fperp}
\\[0.5cm]
G_{\perp}^j
& = & 
\frac{\sqrt{1-\mu^2}}{k_{\perp}}\left[b\left(k_yP_{x\|}^j-k_xP_{y\|}^j\right)-b^{\ast}\left(k_yP_{\|x}^j-k_xP_{\|y}^j\right)\right]
\label{eq:gperp}
\\[0.5cm]
H_{\perp}^j & = & \left(1-\mu^2\right)P_{\|\|}^j
\label{eq:hperp}
\end{eqnarray}
The coefficient (\ref{eq:dperp}), one of the main results of this
paper and presented here in this general form
for the first time, enables one to calculate the perpendicular
diffusion coefficient (\ref{eq:kappaperp}) for an arbitrary turbulence
geometry. Moreover, it allows to evaluate $\kappa_{\perp}$ for a
turbulence consisting of transverse wave modes
with dispersion relations depending arbitrarily on wavevector. 

Whether for numerical or analytical calculations, a further
treatment of its auxiliary functions (\ref{eq:fperp}) to
(\ref{eq:hperp}) requires a certain representation for $P_{\alpha\beta}^j$. 
In view of the structures of equations (\ref{eq:fperp}) to (\ref{eq:hperp}), it is
clear that different representations for $P_{\alpha\beta}^j$ will
alter the underlying mathematical and physical structure of
$D_{\perp}$ and, therefore, $\kappa_{\perp}$.

Here, a representation is chosen commonly used in
the literature. Following, e.g., Lerche \& Schlickeiser (\cite{lersch01}),
the nine components of $P_{\alpha\beta}^j$ can be expressed as 
\begin{equation}
P_{\alpha\beta}^j(k_{\perp},k_{\|})=A^j(k_{\perp},k_{\|})\Biggl[\delta_{\alpha\beta}-\frac{k_{\alpha} k_{\beta}}{k^2}+\imath\,\sigma^j(k_{\perp},k_{\|})\epsilon_{\alpha\beta\nu}\frac{k_{\nu}}{k}\Biggr]
\label{eq:magcorr}
\end{equation}  
where the real quantity $\sigma^j$ denotes the magnetic helicity,
$\delta_{\alpha\beta}$ is Kronecker's delta and
$\epsilon_{\alpha\beta\nu}$ is the Levi-Civita tensor, and $A^j$ is
the wave power spectrum. With the representation (\ref{eq:magcorr}) for $P_{\alpha\beta}^j$,
one can now proceed with the evaluation of equation
(\ref{eq:dperp}). For this, the appropriate components of
(\ref{eq:magcorr}) are substituted into equations (\ref{eq:fperp}),
(\ref{eq:gperp}) and (\ref{eq:hperp}). Making use of equation
(\ref{eq:perpabini}), one arrives at
\begin{eqnarray}
\frac{F_{\perp}^j}{A^j}
& = &
\frac{(a\,k_{\perp}-b\,k_{\|})^2}{k^2}+b^2
+2\frac{|\omega_j|}{v\,k^2}\left(1-\frac{\omega_{j,R}}{|\omega_j|}\right)
\Biggl[\frac{n}{W}k_{\perp}\sqrt{1-\mu^2}+2\mu k_{\|}\Biggr]
\label{eq:fperph}
\\[0.5cm]
\frac{G_{\perp}^j}{A^j}
& = &
-2\imath\sigma^j\sqrt{1-\mu^2}\frac{k_{\perp}}{k}\Biggl[b-\frac{|\omega_j|k_{\|}}{v\,k^2}\left(1-\frac{\omega_{j,R}}{|\omega_j|}\right)\Biggr]
\label{eq:gperph}
\\[0.5cm]
\frac{H_{\perp}^j}{A^j}
& = &
(1-\mu^2)\frac{k_{\perp}^2}{k^2}
\label{eq:hperph}
\end{eqnarray}
where $a$ and $b$ are now real functions, i.e.
\begin{equation}
a=\frac{n}{W}\sqrt{1-\mu^2}-\frac{|\omega_j|k_{\perp}}{v\,k^2}
\,\,\,;\hspace*{0.3cm}
b=\frac{|\omega_j|k_{\|}}{v\,k^2}-\mu
\label{eq:perpab}
\end{equation}
with $|\omega_j|^2=\omega_j\omega_j^{\ast}$. The magnetic helicity
$\sigma^j$ enters the coefficient $D_{\perp}$ only via
$G_{\perp}^j$, which reveals a purely imaginary character,
whereas $F_{\perp}^j$ and $H_{\perp}^j$ are real fields. An evaluation
of equation (\ref{eq:dperp}) involves, therefore, only the real part
of the resonance function (\ref{eq:respwg}), i.e.
\begin{equation}
\Re{\cal R}_{j}=-\frac{\Gamma_j}{\Gamma_j^2+(k_{\|}v_{\|}-\omega_{j,R}+n\Omega)^2}
\label{eq:respwg2}
\end{equation}
which is a positive-definite entry in the net diffusion coefficient
$\kappa_{\perp}$, since $\Gamma_j\leq 0$. 

\section{Diffusion Coefficients for Slab Geometry}
\label{sec:slab}
Although is has been found that the solar wind turbulence is often
dominated by its 2-D modes and has only a small fraction (say $\sim
20\%$) of its energy in the slab contribution (e.g., Matthaeus et
al. \cite{mat90}; Bieber et al. \cite{bibetal94}), it is nevertheless
instructive and desirable to have a solid treatment of quasilinear
perpendicular diffusion in pure slab geometry. For slab
geometry, the wave power spectrum can be given by
\begin{equation}
A^j=g^j(k_{\|})\frac{\delta(k_{\perp})}{k_{\perp}}
\label{eq:powerspec}
\end{equation}
To examine $D_{\perp}$ for the slab geometry limit, the following
approximation process is applied to equation (\ref{eq:dperp}): the
argument $W=k_{\perp}v_{\perp}\sqrt{1-\mu^2}$ of all Bessel functions
is assumed to be much less than unity, since $k_{\perp}\to 0$. Then,
$k_{\perp}$ is small compared to $k_{\|}$, but not equal to zero. This
leads to a ``quasi''-slab model. Finally, the limit $k_{\perp}=0$ is
considered. 

\subsection{Fokker-Planck Coefficient for Slab Geometry}
To start with the approximation, it is convenient to rewrite in
equation (\ref{eq:dperp}) the $n$-summation. Introducing, for
illustrative purposes, an arbitrary function $I(n)$, one gets
\begin{equation}
\sum\limits_{n=-\infty}^{\infty}I(n)=I(0)+\sum\limits_{n=1}^{\infty}[I(-n)+I(n)]=\sum\limits_{r=\pm 1}\sum\limits_{n=0}^{\infty}I(rn)\neq\sum\limits_{n=0}\sum\limits_{r=\pm 1}^{\infty}I(rn)
\label{eq:sumaux}
\end{equation}
where the last step indicates that the sequence of summation may not be switched. Based on equation (\ref{eq:sumaux}), the coefficient (\ref{eq:dperp}) can be written as
\begin{eqnarray}
D_{\perp} & = & \frac{v^2}{B_0^2}
\sum\limits_{j=\pm 1}^{}
\sum\limits_{r=\pm 1}^{}
\sum\limits_{n=0}^{\infty}
\int\limits_{}^{}d^3 k {\cal R}_{j}(rn)
\label{eq:dperp2}
\\
&&\times
\Biggl[J_{n}^2(W)F_{\perp}^j(rn)+\imath J_{n}(W)J_{n}^{\prime}(W)G_{\perp}^j(rn)+\bigl[J_{n}^{\prime}(W)\bigr]^2H_{\perp}^j(rn)\Biggr]
\nonumber
\end{eqnarray}
The argument occurring at the resonance function and the auxiliary functions indicates that one has to change in the corresponding equations the quantity $n$ to $rn$. The Bessel functions are not affected by the new sum index $r$. For slab geometry, the asymptotic
\begin{equation}
J_n(W)\simeq \frac{1}{\Gamma(1+n)}\left(\frac{W}{2}\right)^n
\label{eq:besselapp}
\end{equation}
is used, representing the Bessel function for small arguments $W\to 0$, with $\Gamma(x)$ being the Gamma function. Substitution of expression (\ref{eq:besselapp}) into the coefficient (\ref{eq:dperp2}) results in
\begin{equation}
D_{\perp}= \frac{v^2}{B_0^2}
\sum\limits_{j=\pm 1}^{}
\sum\limits_{r=\pm 1}^{}
\sum\limits_{n=0}^{\infty}
\int\limits_{}^{}d^3k
\frac{{\cal R}_{j}(rn)}{\Gamma^2(1+n)}\left(\frac{W}{2}\right)^{2n}
\Biggl[F_{\perp}^j(rn)+\imath G_{\perp}^j(rn)\frac{n}{W}+H_{\perp}^j\frac{n^2}{W^2}\Biggr]
\label{eq:quasislabperp}
\end{equation}
A further treatment of equation (\ref{eq:quasislabperp}) requires a closer inspection of equations (\ref{eq:fperph}), (\ref{eq:gperph}) and (\ref{eq:hperph}). They can be represented as polynomials of different orders in the ratio $n/W$: the field $F_{\perp}^j$ is a second-order polynomial in $n/W$, whereas $G_{\perp}^j$ and $H_{\perp}^j$ are polynomials of zeroth-order. The term in the brackets of equation (\ref{eq:quasislabperp}) then yields
\begin{equation}
F_{\perp}^j(rn)+\imath G_{\perp}^j(rn)\frac{n}{W}+H_{\perp}^j\frac{n^2}{W^2}
=
A^j(k_{\perp},k_{\|})\Biggl[\alpha_1\left(\frac{n}{W}\right)^2+\alpha_2\left(\frac{n}{W}\right)+\alpha_3\Biggr]
\label{eq:poly}
\end{equation}
where the coefficients $\alpha_1$, $\alpha_2$ and $\alpha_3$ read
\begin{eqnarray}
\alpha_1 & = & 2(1-\mu^2)\frac{k_{\perp}^2}{k^2}
\label{eq:alp1}
\\
\alpha_2 & = & \sqrt{2\alpha_1}\left[\frac{(k_{\|}v_{\|}-2\omega_{j,R}+|\omega_j|)r}{vk}+\sigma^j\left(\frac{\omega_{j,R}k_{\|}}{vk^2}-\mu\right)\right]
\label{eq:alp2}
\\
\alpha_3 & = & \left(\frac{|\omega_{j}|k_{\|}}{vk^2}+\mu\right)^2-\frac{4\mu\omega_{j,R}k_{\|}}{vk^2}+\frac{(k_{\|}v_{\|}-\omega_{j,R})^2}{v^2k^2}
\label{eq:alp3}
\end{eqnarray}
With equations (\ref{eq:poly}) to (\ref{eq:alp3}), the perpendicular diffusion coefficient (\ref{eq:quasislabperp}) can be rearranged to become $D_{\perp}=D_{\perp,1}+D_{\perp,2}+D_{\perp,3}$, where
\begin{eqnarray}
D_{\perp,1} & = & \frac{v^2}{4B_0^2}\sum\limits_{j=\pm 1}^{}
\sum\limits_{r=\pm 1}^{}
\int\limits_{}^{}dk^3\alpha_1A^j(k_{\perp},k_{\|})\sum\limits_{n=0}^{\infty}n^2
\frac{{\cal R}_{j}(rn)}{\Gamma^2(1+n)}\left(\frac{W}{2}\right)^{2n-2}
\label{eq:d1}
\\
D_{\perp,2} & = & \frac{v^2}{2B_0^2}\sum\limits_{j=\pm 1}^{}
\sum\limits_{r=\pm 1}^{}
\int\limits_{}^{}dk^3\alpha_2A^j(k_{\perp},k_{\|})\sum\limits_{n=0}^{\infty}n
\frac{{\cal R}_{j}(rn)}{\Gamma^2(1+n)}\left(\frac{W}{2}\right)^{2n-1}
\label{eq:d2}
\\
D_{\perp,3} & = & \frac{v^2}{B_0^2}\sum\limits_{j=\pm 1}^{}
\sum\limits_{r=\pm 1}^{}
\int\limits_{}^{}dk^3\alpha_3A^j(k_{\perp},k_{\|})\sum\limits_{n=0}^{\infty}
\frac{{\cal R}_{j}(rn)}{\Gamma^2(1+n)}\left(\frac{W}{2}\right)^{2n}
\label{eq:d3}
\end{eqnarray}
For the limit of pure slab geometry, the $n$-summations in the integrals are subjected to a closer inspection. The $n=0$ contribution vanishes in any case for $D_{\perp,1}$ and $D_{\perp,2}$. For $W\propto k_{\perp}=0$, it is clear that only  $n=1$ can contribute to the sum appearing in $D_{\perp,1}$. Concerning $D_{\perp,2}$, all contributions vanish, and one obtains
\begin{equation}
\sum\limits_{n=0}^{\infty}n^2\frac{{\cal R}_{j}(rn)}{\Gamma^2(1+n)}\left(\frac{W}{2}\right)^{2n-2}
={\cal R}_{j}(r)
\,\,\,;\hspace*{0.3cm}
\sum\limits_{n=0}^{\infty}n\frac{{\cal R}_{j}(rn)}{\Gamma^2(1+n)}\left(\frac{W}{2}\right)^{2n-1}
=0
\label{eq:rel1}
\end{equation}
The evaluation of the sum in equation (\ref{eq:d3}) leads to
\begin{equation}
\sum\limits_{n=0}^{\infty}
\frac{{\cal R}_{j}(rn)}{\Gamma^2(1+n)}\left(\frac{W}{2}\right)^{2n}={\cal R}_{j}(n=0)
\label{eq:ref2}
\end{equation}
The second relation in equation (\ref{eq:rel1}) implies
$D_{\perp,2}=0$. Consequently, the magnetic helicity $\sigma^j$ does
not affect QLT perpendicular diffusion of particles in a slab plasma
wave turbulence. Substitution of the left-handed expression of
equation (\ref{eq:rel1}) into (\ref{eq:d1}), and making use of
(\ref{eq:alp1}) and the wave power spectrum (\ref{eq:powerspec}), yields
\begin{equation}
D_{\perp,1}^{slab} = \frac{\pi v^2}{B_0^2}(1-\mu^2)\sum\limits_{j=\pm 1}^{}
\sum\limits_{r=\pm 1}^{}
\int\limits_{-\infty}^{\infty}dk_{\|}g^j(k_{\|}){\cal R}_{j}(r)
\int\limits_{0}^{\infty}dk_{\perp}\frac{\delta(k_{\perp})k_{\perp}^2}{k^2_{\|}+k_{\perp}^2}=0
\end{equation}
Apparently, $D_{\perp,3}$ is the only nonvanishing contribution for
slab geometry. Rearranging some terms in $\alpha_3$, equation
(\ref{eq:alp3}), one arrives at
\begin{eqnarray}
D_{\perp}^{slab} = D_{\perp,3}^{slab} & = & \frac{2\pi v^2}{B_0^2}\sum\limits_{j=\pm 1}^{}
\int\limits_{-\infty}^{\infty}dk_{\|}g^j(k_{\|}){\cal R}_{j}(n=0)
\label{eq:fpcgen}
\\
&&\times
\left[2\frac{(k_{\|}v\mu-\omega_{j,R})^2}{v^2k_{\|}^2}+\frac{\Gamma_j^2(k_{\|})}{v^2k_{\|}^2}+2\frac{\mu}{vk_{\|}}\left(|\omega_j|-\omega_{j,R}\right)\right]
\nonumber
\end{eqnarray}
Equation (\ref{eq:fpcgen}) is the general FPC for perpendicular
diffusion of particles in a slab turbulence consisting of transverse,
damped plasma waves. The wave power spectrum $g^j$, the real frequency
$\omega_{j,R}$ of the plasma wave mode and the corresponding
dissipation rate $\Gamma_j$ depend arbitrarily on $k_{\|}$. Equation
(\ref{eq:fpcgen}) allows to derive a general diffusion coefficient 
$\kappa_{\perp}^{slab}$ not only for the plasma wave viewpoint, but
also for the so-called dynamical magnetic turbulence
approach. Perpendicular diffusion coefficients for slab geometry and
for these two approaches are derived in the next two sections

\subsection{Plasma Wave Turbulence Approach}
\label{sec:pwt}
After having considered the slab limit of the general Fokker-Planck
coefficient (\ref{eq:dperp}) in 
some detail, the corresponding diffusion coefficient
$\kappa_{\perp}^{slab}$ is derived in this section. For this, the pitch-angle
integration as given by equation (\ref{eq:kappaperp}) is performed and
then, if necessary, the integration in wavenumber space. Substitution
of equation (\ref{eq:fpcgen}) into (\ref{eq:kappaperp}) results in
\begin{equation}
\kappa_{\perp}^{slab}=\frac{1}{2}\int\limits_{-1}^{+1}d\mu
D_{\perp}^{slab}=-\frac{\pi v^2}{B_0^2}\sum\limits_{j=\pm
  1}^{}\int\limits_{-\infty}^{\infty}dk_{\|}g^j\Gamma_j\left[I_1+I_2+I_3\right]
\label{eq:kappa2}
\end{equation}
where the functions $I_1$, $I_2$ and $I_3$ are pitch-angle
integrals. Furthermore, the resonance function (\ref{eq:respwg2}) has
been used. The pitch-angle integrals can be solved analytically, yielding 
\begin{equation}
I_1=\frac{2}{v^3k_{\|}^3}\int\limits_{-B+}^{+B_-}dx\frac{x^2}{\Gamma_j^2+x^2}=\frac{4}{v^2k_{\|}^2}-\frac{2\Gamma_j}{v^3k_{\|}^3}\left[\arctan\left(\frac{B_+}{\Gamma_j}\right)+\arctan\left(\frac{B_-}{\Gamma_j}\right)\right]
\label{eq:int1}
\end{equation}
\begin{equation}
I_2=\frac{\Gamma_j^2}{v^3k_{\|}^3}\int\limits_{-B+}^{+B_-}dx\frac{1}{\Gamma_j^2+x^2}=\frac{\Gamma_j}{v^3k_{\|}^3}\left[\arctan\left(\frac{B_+}{\Gamma_j}\right)+\arctan\left(\frac{B_-}{\Gamma_j}\right)\right]
\label{eq:int2}
\end{equation}
\begin{eqnarray}
I_3=\frac{2(|\omega_j|-\omega_{j,R})}{v^3k_{\|}^3}\int\limits_{-B+}^{+B_-}dx\frac{x+\omega_{j,R}}{\Gamma_j^2+x^2} & = & \frac{(|\omega_j|-\omega_{j,R})}{v^3k_{\|}^3}\Biggl\{\ln(B_-^2+\Gamma_j^2)-\ln(B_+^2+\Gamma_j^2)
\nonumber
\\
&&+\frac{\omega_{j,R}}{\Gamma_j}\left[
\arctan\left(\frac{B_+}{\Gamma_j}\right)+\arctan\left(\frac{B_-}{\Gamma_j}\right)\right]\Biggr\}
\label{eq:int3}
\end{eqnarray}
where the integration boundaries are given by
$B_{\pm}=k_{\|}v\pm\omega_{j,R}$. Concerning the term
$|\omega_j|-\omega_{j,R}=\omega_{j,R}\left[\left(1+\Gamma_j^2/\omega_{j,R}^2\right)^{1/2}-1\right]$
in equation (\ref{eq:int3}), it is obvious that $I_3$ becomes
negligible for the condition $|\Gamma_j|\ll|\omega_{j,R}|$. This is an
often used assumption within quasilinear theory to derive the
dissipation rates of plasma wave modes, and the approximation $I_3=0$
is considered as reasonable for further progress.

With equations (\ref{eq:int1}) and (\ref{eq:int2}), the quasilinear
perpendicular diffusion coefficient for a plasma wave turbulence with
slab geometry reads
\begin{equation}
\kappa_{\perp}^{slab}=\frac{2\pi}{B_0^2}\sum\limits_{j=\pm
  1}^{}\int\limits_{0}^{\infty}dk_{\|}g^jk_{\|}^{-2}\Gamma_j\left\{4-\frac{\Gamma_j}{vk_{\|}}\left[\arctan\left(\frac{B_+}{\Gamma_j}\right)+\arctan\left(\frac{B_-}{\Gamma_j}\right)\right]\right\}
\label{eq:kappafin}
\end{equation}
where it was assumed that $g^j(k_{\|})=g^j(|k_{\|}|)$,
$\Gamma_j(k_{\|})=\Gamma_j(|k_{\|}|)$ and
$\omega_{j,R}(k_{\|})=\omega_{j,R}(|k_{\|}|)$. Furthermore, the fact was used
that $\arctan(x)$ is an odd function in $x$. As a consequence of
this, the dissipation rate $\Gamma_j$ is now given as a positive
quantity, meaning that $\Gamma_j\geq 0$ in equation
(\ref{eq:kappafin}).

Further evaluation of the wavenumber integral in equation
(\ref{eq:kappafin}) requires specific wavenumber variations of $g^j$, $\omega_{j,R}$ and
$\Gamma_j$. Especially the dissipation rate $\Gamma_j$ can usually be
given as a quite complicated function not only in $k_{\|}$, but also
in the so-called plasma $\beta_{p,e}=8\pi n_{p,e}k_B
T_{p,e}/B_0^2$. In the latter, the subscripts refer to protons and electrons as plasma
constituents having the corresponding temperatures $T_{p,e}$ and the
number densities $n_{p,e}$. The temperatures of the plasma electrons
and protons can differ quite significantly, giving rise to
instabilities (e.g., Gary \cite{gar93}). It is well known that such a
plasma configuration with different temperatures of the plasma
components can then result in a variety of wave modes with
substantially different $\omega_{j,R}$ and dissipation rates
$\Gamma_j$. 

For instance, consider the Alfv\'en wave mode with real frequency
$\omega_{j,R}=jv_Ak_{\|}$. A typical representation for the damping
rate of this wave mode can be derived by using a quasilinear Vlasov
theory code (e.g., Gary \cite{gar93}). Doing this, one obtains
\begin{equation}
\Gamma_j(k_{\|})=0.60\beta_p^{0.36}\Omega_p\left(\frac{k_{\|}c}{\omega_p}\right)^{1.54\beta_p^{0.03}}\exp\left(-\frac{0.32\omega_p^2}{\beta_p^{0.65}k_{\|}^2c^2}\right)
\label{eq:dampingalf}
\end{equation}   
where $\omega_{p}$ and $\Omega_p$ are the plasma frequency and the
plasma proton gyrofrequency, respectively  (Stawicki et
al. \cite{sta2}). In view of (\ref{eq:dampingalf}), it is clear that the
wavenumber integration in equation (\ref{eq:kappafin}) has to
be solved numerically. This is beyond the scope of this paper, and,
for illustrative purposes, a simplified version of the damping rate
(\ref{eq:dampingalf}) is demanded. For this, the exponential
expression in Eq. (\ref{eq:dampingalf}) is roughly
neglected. Furthermore, it is assumed that $\beta_p$ is such that the exponent of the
power law is equal to unity. The two crude simplifications lead to
\begin{equation}
\Gamma_j(k_{\|})=\Gamma_{j,0}v _A k_{\|}
\end{equation}
where $\Gamma_{j,0}$ is a constant. Substitution into (\ref{eq:kappafin}) leads to
\begin{equation}
\kappa_{\perp}^{slab}=\frac{4\pi v\eta}{B_0^2}\left\{4-\eta\left[\arctan\left(\frac{1+v_A/v}{\eta}\right)+\arctan\left(\frac{1-v_A/v}{\eta}\right)\right]\right\}
\int\limits_{0}^{\infty}dk_{\|}\frac{g(k_{\|})}{k_{\|}}
\label{eq:kappaalf2}
\end{equation}
where it was assumed that $\Gamma_{+,0}=\Gamma_{-,0}=\Gamma_0$ and $g^+=g^-=g$. Furthermore, the quantity $\eta=\Gamma_0v_A/v$ has been introduced. Apparently, the expression in the curly brackets of (\ref{eq:kappafin}) can be shifted in front of the wavenumber integral. It has to be stressed that this is a consequence of the crude approximations made to simplify the dissipation rate (\ref{eq:dampingalf}): $\Gamma_j$ is linear in $k_{\|}$ and, therefore, wipes out the wavenumber dependence of $\omega_{j,R}$. 

Taking into account that $v_A/v\ll 1$ and neglecting the corresponding terms in equation (\ref{eq:kappaalf2}), one obtains
\begin{equation}
\kappa_{\perp}^{slab}=\frac{8\pi v\eta}{B_0^2}\left[2-\eta\arctan\left(\eta^{-1}\right)\right]
\int\limits_{0}^{\infty}dk_{\|}\frac{g(k_{\|})}{k_{\|}}
\label{eq:kappaalf3}
\end{equation}
A closer inspection of equation (\ref{eq:kappaalf3}), and its general
version, equation (\ref{eq:kappafin}), results in the following important observation:
$\kappa_{\perp}^{slab}$ vanishes for a dissipationless plasma wave turbulence,
i.e. $\Gamma_j=0$. Consequently, for a static turbulence, charged
particles can not move in perpendicular direction and are trapped to a
line of force of the background magnetic field.


\subsubsection{Dissipationless Alv\'enic Turbulence}
\begin{equation}
{\cal R}_{j}(n=0)=\pi\delta(k_{\|}v\mu-jv_Ak_{\|})
\label{eq:diracapp}
\end{equation}
where the Alfv\'en dispersion relations $\omega_{j,R}=jv_Ak_{\|}$ was used.
For $\Gamma_j=0$, the second and third term in the brackets of equation
(\ref{eq:fpcgen}) vanish in any case. The first term remains as
the only one for the following two derivations. 
\\
Derivation I: In order to obtain $\kappa_{\perp}^{slab}$, the
wavenumber integration is first performed, and then the
$\mu$-integration. Making use of equation (\ref{eq:diracapp}), one
obtains for the FPC (\ref{eq:fpcgen}) the expression
\begin{equation}
D_{\perp}^{slab} =\frac{4\pi^2}{B_0^2}\sum\limits_{j=\pm 1}^{}
\int\limits_{-\infty}^{\infty}dk_{\|}g^j(k_{\|})\delta(k_{\|})|v\mu-jv_A|=\frac{4\pi^2}{B_0^2}\sum\limits_{j=\pm 1}^{}g^j(0)|v\mu-jv_A|
\end{equation}
The argument of the wave power spectrum $g^j$ indicates that is has to be evaluated for $k_{\|}=0$. Equation (\ref{eq:kappaperp}) then yields
\begin{equation}
\kappa_{\perp}^{slab}=\frac{1}{2}\int\limits_{-1}^{+1}d\mu D_{\perp}^{slab}=2\pi^2v\frac{[g^+(0)+g^-(0)]}{B_0^2}\int\limits_{0}^{1}d\mu\biggl[|\mu-jv_A/v|+|\mu+jv_A/v|\biggr]
\end{equation}
The integrand can be approximated by $2v_A/v$ and $2\mu$ for the
intervals $0\leq\mu\leq v_A/v$ and $v_A/v\leq\mu\leq 1$, respectively,
and one arrives at
\begin{equation}
\kappa_{\perp}^{slab}=2\pi^2v\frac{[g^+(0)+g^-(0)]}{B_0^2}\left(1+\frac{v_A^2}{v^2}\right)\simeq \frac{\pi v L}{2}\left(\frac{\delta B(0)}{B_0}\right)^2
\label{eq:kappaalf}
\end{equation}
In the last step of equation (\ref{eq:kappaalf}) it was made use of
$g^+(0)+g^-(0)=\delta B^2(0)L/4\pi$, with $L$ being a 
scale length providing for the right normalization. 
The neglect of $v_A^2/v^2\ll 1$
corresponds to a vanishing irregular electric field and, therefore,
results in a purely magnetic turbulence. Equation (\ref{eq:kappaalf})
agrees with earlier quasilinear results interpreted as non-resonant
field line random walk in a magnetostatic turbulence (Jokipii
\cite{jo1}; Forman \cite{for1}). 
\\
Derivation II: The second derivation is based on the fact, that one
can easily switch the order of integration: first, the
$\mu$-integration is performed, and then, if necessary, the
integration over $k_{\|}$. Substitution of the resonance function
(\ref{eq:diracapp}) into Eq. (\ref{eq:fpcgen}) results in
\begin{eqnarray}
D_{\perp}^{slab} & = &
\frac{4\pi^2}{B_0^2}\sum\limits_{j=\pm 1}^{}
\int\limits_{-\infty}^{\infty}dk_{\|}g^j(k_{\|})k_{\|}^{-2}\delta(k_{\|}v\mu-jv_Ak_{\|})(k_{\|}v\mu-jv_Ak_{\|})^2
\nonumber
\\
& = &
\frac{8\pi^2v}{B_0^2}\sum\limits_{j=\pm 1}^{}
\int\limits_{0}^{\infty}dk_{\|}g^j(k_{\|})k_{\|}^{-1}\delta(\mu-jv_A/v)(\mu-jv_A/v)^2
\end{eqnarray}
Applying now the pitch-angle integration, according to
Eq. (\ref{eq:kappaperp}), one obtains
\begin{equation}
\kappa_{\perp}^{slab}=\frac{4\pi^2v}{B_0^2}\sum\limits_{j=\pm 1}^{}
\int\limits_{0}^{\infty}dk_{\|}g^j(k_{\|})k_{\|}^{-1}\int\limits_{-1}^{+1}d\mu\delta(\mu-jv_A/v)(\mu-jv_A/v)^2=0
\label{eq:kastatic}
\end{equation}
Therefore, the perpendicular diffusion coefficient vanishes for a
dissipationless turbulence. As expected, the result
(\ref{eq:kastatic}) then agrees with the general diffusion coefficient
(\ref{eq:kappafin}).


\subsection{Dynamical Magnetic Turbulence Approach}
\label{sec:dmt2}
In this section, derivations of quasilinear perpendicular
diffusion coefficients for the so-called dynamical magnetic turbulence
(DMT) model, introduced by Bieber et al. (\cite{bibetal94}), are presented.
Within the context of DMT, fluctuations are assumed to be purely
magnetic. To take into account the dynamical behavior of such purely magnetic
fluctuations, Bieber et al. (\cite{bibetal94}) defined two models: the
damping as well as the random sweeping model. It is relatively
straightforward to derive $\kappa_{\perp}^{slab}$ on the basis of the more
general FPC (\ref{eq:fpcgen}) for these two models, and both are considered in turn.
The limit of vanishing turbulent electric fields can easily
be achieved by setting $\omega_{j,R}=0$ and $\Gamma_j=0$ in
Eq. (\ref{eq:fpcgen}), initially derived for the
plasma wave viewpoint. The only modification concerns the real part of
the resonance function (\ref{eq:respwg}). Furthermore, since the
concept of a superposition of individual wave modes does not apply
anymore, the $j$-nomenclature is dropped. 

\subsubsection{Damping Model}
For the damping model, Bieber et al. (\cite{bibetal94}) suggested a
dynamical behavior of the turbulent energy being of the form
\begin{equation}
<\delta B({\bf k},t)\delta B^{\ast}({\bf k},t+\xi)>=P({\bf k})\exp(-\nu_c\xi)
\end{equation}
and the resonance function (\ref{eq:respwg}) has to be modified in
this respect. What has to be done is to set $\Gamma_j=0$ in the upper
part of Eq. (\ref{eq:respwg}), and then multiply the exponential
expression with $\exp(-\nu_c\xi)$. The contributions resulting from
the unperturbed particle orbit still hold. Having in mind that one has to take the real part of
equation (\ref{eq:respwg}) (see Eq. (\ref{eq:perpab}) and the comments
following it), the resonance function for the damping model reads
\begin{equation}
\Re{\cal R}=\Re\int\limits_{0}^{\infty}d\xi
\exp\left[-\imath(k_{\|}v_{\|}+n\Omega)\xi-\nu_c\xi\right]= 
\frac{\nu_c}{\nu_{c}^2+(k_{\|}v_{\|}+n\Omega)^2}
\label{eq:dmtres}
\end{equation}
Following Bieber et al. (\cite{bibetal94}), the rate for turbulent
decorrelation, $\nu_c$, is assumed to be
\begin{equation}
\nu_c=\alpha_m v_A k_{\|}
\label{eq:nudmt}
\end{equation}
where the parameter $0\leq \alpha_m\leq 1$ allows adjustment of the
strength of the dynamical effects. The case $\alpha_m=0$ then
represents the magnetostatic limit, $\alpha_m=1$ describes a strongly
dynamical magnetic turbulence. With the decorrelation rate (\ref{eq:nudmt}) and the resonance
function (\ref{eq:dmtres}), one readily obtains for
Eq. (\ref{eq:fpcgen}) the expression
\begin{equation}
D_{\perp}^{slab} = \frac{4\pi v^2}{B_0^2}\int\limits_{-\infty}^{\infty}dk_{\|}g(k_{\|})\mu^2{\cal R}(n=0)=\frac{8\pi v\zeta}{B_0^2}\frac{\mu^2}{\zeta^2+\mu^2\zeta^2}\int\limits_{0}^{\infty}dk_{\|}\frac{g(k_{\|})}{k_{\|}}
\end{equation}
where $\zeta=\alpha_m v_A/v$. Furthermore,
$g(k_{\|})=g(|k_{\|}|)$. With equation (\ref{eq:kappaperp}), one obtains
\begin{equation}
\kappa_{\perp}^{slab}=\frac{1}{2}\int\limits_{-1}^{+1}d\mu D_{\perp}^{slab}=\frac{8\pi v\zeta}{B_0^2}\int\limits_{0}^{\infty}dk_{\|}\frac{g(k_{\|})}{k_{\|}}\int\limits_{0}^{1}d\mu\frac{\mu^2}{1+\mu^2\zeta^2}
\end{equation}
where the $\mu$-integration can be carried out analytically, yielding
\begin{equation}
\kappa_{\perp}^{slab}=\frac{8\pi v\zeta}{B_0^2}\biggl[1-\zeta\arctan\left(\zeta^{-1}\right)\biggr]\int\limits_{0}^{\infty}dk_{\|}\frac{g(k_{\|})}{k_{\|}}
\label{eq:kappadmt}
\end{equation}
Obviously, for the limit $\alpha_m=0$ (magnetostatic limit),
 $\kappa_{\perp}^{slab}$ vanishes, implying that the particle remains
 tied to the background magnetic field. Shalchi \& Schlieiser
 (\cite{shalchischlickeiser2004}) came to the same conclusion.

\subsubsection{Random Sweeping Model}
For the random sweeping model,  Bieber et al. (\cite{bibetal94}) used a
Gaussian dependence for the turbulence decay, i.e.
\begin{equation}
<\delta B({\bf k},t)\delta B^{\ast}({\bf k},t+\xi)>=P({\bf k})\exp(-\nu_c^2\xi^2)
\end{equation}
This changes the governing resonance function (\ref{eq:respwg}) significantly, and one obtains
\begin{equation}
\Re{\cal R}=\int\limits_{0}^{\infty}d\xi
\exp\left[-\imath(k_{\|}v_{\|}+n\Omega)\xi-\nu_c^2\xi^2\right]=\frac{\sqrt{\pi}}{2\nu_c}\exp\left(-\frac{(k_{\|}v_{\|}+n\Omega_{\alpha})^2}{4\nu_c^2}\right)
\label{eq:random}
\end{equation}
Making use of equation (\ref{eq:fpcgen}) and setting $\omega_j=\Gamma_j=0$, one arrives at 
\begin{equation}
\kappa_{\perp}^{slab}=\frac{1}{2}\int\limits_{-1}^{+1}d\mu D_{\perp}^{slab}=\frac{4\pi^{3/2}v^2}{B_0^2}\int\limits_{0}^{\infty}dk_{\|}\frac{g(k_{\|})}{\nu_c(k_{\|})}\int\limits_{0}^{1}d\mu\mu^2\exp\left(-\frac{k_{\|}^2v^2\mu^2}{4\nu_c(k_{\|})}\right)
\end{equation}
The pitch-angle integral can be solved again analytically, resulting in
\begin{equation}
\kappa_{\perp}^{slab}=\frac{8\pi^{3/2}}{B_0^2v}\int\limits_{0}^{\infty}dk_{\|}\frac{g(k_{\|})}{k_{\|}^2}\nu_c^2(k_{\|})\left[\sqrt{\pi}\mbox{Erf}\left(\frac{k_{\|}v}{2\nu_{c}(k_{\|})}\right)-\frac{k_{\|}v}{\nu_{c}(k_{\|})}\exp\left(-\frac{k_{\|}^2v^2}{4\nu_{c}^2(k_{\|})}\right)\right]
\label{eq:kapparan}
\end{equation}
where $\mbox{Erf}(x)$ is the Error function. So far, equation (\ref{eq:kapparan}) is the general representation of the perpendicular diffusion coefficient in a slab turbulence obeying the random sweeping model. Upon using for the decorrelation rate the same dependence on $k_{\|}$ as for the damping model, i.e. $\nu_c=\alpha_mv_Ak_{\|}$, one obtains
\begin{equation}
\kappa_{\perp}^{slab}=\frac{8\pi^{3/2}v\zeta^2}{B_0^2}\left[\sqrt{\pi}\mbox{Erf}\left(\frac{\zeta^{-1}}{2}\right)-\frac{\zeta^{-1}}{2}\exp\left(-\frac{\zeta^{-1}}{4}\right)\right]\int\limits_{0}^{\infty}dk_{\|}g(k_{\|})
\end{equation}
Again, $\kappa_{\perp}^{slab}$ vanishes for the magnetostatic limit
$\alpha_m\to 0$. 

\section{Diffusion Coefficient for 2D Geometry} 
\label{sec:2d}
The previous section offered detailed insight into the slab limit of
the general FPC (\ref{eq:dperp}) and its associated transport
parameter (\ref{eq:kappaperp}). Perpendicular diffusion coefficients
are presented for the slab limit for both the plasma wave viewpoint and the
dynamical magnetic turbulence approach. In this section, the
evaluation of the perpendicular diffusion coefficient for the
2D contribution of the turbulence is presented. For simplicity, the
calculations are restricted to the damping model of the DMT
description (see also Section \ref{sec:dmt2}), and the more general
case of a plasma wave turbulence is left as an exercise to the
interested reader. The calculations presented in this section
generalize the derivation presented by Shalchi \& Schlickeiser (\cite{shalchischlickeiser2004}), since
their approach is restricted to a simplified magnetic power spectrum
and turbulence decorrelation rate.

\begin{figure}
\includegraphics[width=9cm]{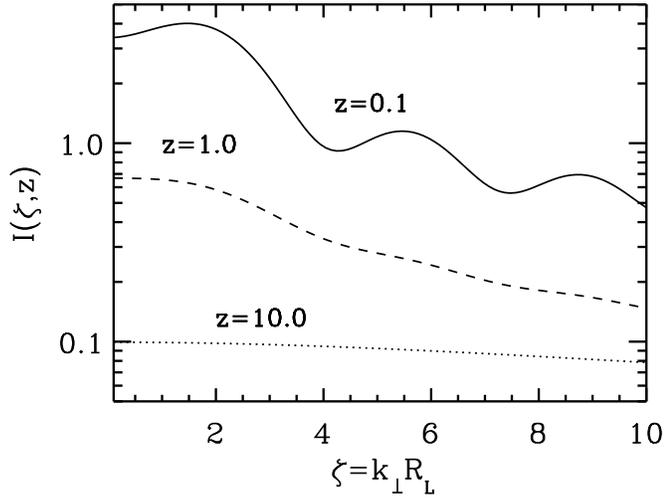}
\caption{Numerical solutions of Eq. (\ref{eq:funcI}) representing
  the behavior of $I(\zeta,z)$ for three different values of $z$}
\label{fig:fig1}
\end{figure}

For 2D turbulence, wavevectors are
perpendicular to the mean magnetic field, and the wave power spectrum
can be given by
\begin{equation}
A^j(k_{\perp},k_{\|})=g(k_{\perp})\frac{\delta(k_{\|})}{k_{\perp}}
\label{eq:power2d}
\end{equation}
Upon substituting the expression (\ref{eq:power2d}) into equation
(\ref{eq:dperp}) and making use of the Bessel function identities (\ref{eq:besselid}),
the FPC for 2D geometry can be cast into the form
\begin{equation}
D_{\perp}^{2D}=\frac{2\pi v^2}{B_0^2}\sum\limits_{n=-\infty}^{\infty}
\int\limits_{0}^{\infty}dk_{\perp}\frac{g(k_{\perp})\nu_c(k_{\perp})}{\nu_{c}^2(k_{\perp})+n^2\Omega^2}
\left[\frac{(1-\mu^2)}{2}\left(J_{n+1}^2(W)+J_{n-1}^2(W)\right)+\mu^2J_n^2(W)\right]
\label{eq:nonres}
\end{equation}
where $\sigma=0$ was assumed. 
The corresponding resonance function has already been inserted with
$\nu_c(k_{\perp})$ being the decorrelation rate in normal direction
(cf. Eq. (\ref{eq:dmtres})). According to equation
(\ref{eq:kappaperp}), one has to perform the
$\mu$-integration to obtain the corresponding diffusion coefficient
$\kappa_{\perp}^{2D}$. The integration can still be carried out
analytically, and the detailed calculations are presented in Appendix
\ref{sec:app}. There, it is shown that the diffusion coefficient for
2D geometry reads
\begin{equation}
\kappa_{\perp}^{2D}=\frac{\pi vR_L}{B_0^2}\int\limits_{0}^{\infty}dk_{\perp}g(k_{\perp})I(\zeta,z)
\label{eq:kappa2d}
\end{equation}
with
\begin{equation}
I(\zeta,z)=\int\limits_{0}^{\pi/2}d\theta\frac{\cosh(2\theta z)}{\zeta^{3}\sinh(z\pi)\cos\theta}\biggl[(1-2\zeta^2\cos(2\theta))\sin(2\zeta\cos\theta)-2\zeta\cos\theta\cos(2\zeta\cos\theta)\biggr]
\label{eq:funcI}
\end{equation}
where the abbreviations $\zeta=k_{\perp}R_L$ and $z=\nu_c/\Omega$ are
introduced. Furthermore, $R_L=v/\Omega$ is the Larmor radius.
Equation (\ref{eq:kappa2d}) is valid for a power spectrum and a
decorrelation rate varying arbitrarily in wavenumber $k_{\perp}$.
The integral representation (\ref{eq:funcI}) results from the
$\mu$-integration and has to be evaluated for further
progress. Unfortunately, an analytical solution for this integral does
not exist, and any progress requires numerical treatment. In order to obtain some insight into the
behavior of the function $I(\zeta,z)$, Figure \ref{fig:fig1} shows numerical
solutions of Eq. (\ref{eq:funcI}) as function of $\zeta$ for three
different values of $z$. For illustrative purposes, $z$ is assumed to
be constant in $k_{\perp}$. 

The limit $\zeta\ll 1$, implying $R_L\ll k_{\perp}^{-1}$, leads to an instructive, analytical
solution for equation (\ref{eq:funcI}). To show this,
small argument approximations for the circular functions are used, i.e.
\begin{equation}
\cos(2\zeta\cos\theta)\simeq 1;\,\,\,\,\,
\sin(2\zeta\cos\theta)\simeq 2\zeta\cos\theta
\end{equation}
and inserted into $I(\zeta,z)$. Partial integration then results in
\begin{equation}
I(\zeta\ll 1,z)=\frac{z}{1+z^2}
\end{equation}
Consequently, one obtains
\begin{equation}
\kappa_{\perp}^{2D}=\frac{2\pi
v}{B_0^2}R_L\int\limits_{0}^{\infty}dk_{\perp}g(k_{\perp})\frac{(\tau_c\Omega)}{1+(\tau_c\Omega)^2}
\label{eq:trapafin}
\end{equation}
for the limit $R_L\ll k_{\perp}^{-1}$, where the relation
$\nu_c=\tau_{c}^{-1}$ has been used. An eyecatching feature of equation (\ref{eq:trapafin}) is the term
including the dimensionless product $\tau_c\Omega$. It is formally the
same as those derived by Forman et al. (\cite{for1}) for field
line random walk in a slab geometry and, more recently, by Bieber \& Matthaeus (\cite{biemat97}).

For the magnetostatic limit, $\nu_c\to 0$ or $z=0$, it can be shown easily
that $\kappa_{\perp}^{2D}\to\infty$, by using small argument
approximations for the hyperbolic functions. 
However, for slab geometry, the diffusion coefficients
approach zero for $\nu_c\to 0$ (see section \ref{sec:slab}).

\section{Numerical Calculations and Conclusions}
\label{sec:num}
To demonstrate the potential and flexibility provided with the new
perpendicular diffusion coefficients for slab and 2D geometry, typical
parameter for heliospheric conditions are applied and the remaining
wavenumber integrations are solved numerically. The numerical
calculations are performed for $\kappa_{\perp}^{slab}$ and
$\kappa_{\perp}^{2D}$ as given by equations  (\ref{eq:kappadmt}) and
(\ref{eq:kappa2d}), respectively, for the damping model of the DMT
approach.

For the numerical computation, the power spectra
\begin{equation}
g(k_{\|})=C(q)\lambda_{slab}\delta B^2_{slab}(1+k_{\|}^2\lambda_{slab}^2)^{-q}
\label{eq:pows}
\end{equation} 
and
\begin{equation}
g(k_{\perp})=C(q)\lambda_{2D}\delta B^2_{2D}(1+k_{\perp}^2\lambda_{2D}^2)^{-q}
\label{eq:pow2d}
\end{equation} 
are used for the slab and 2D contribution, respectively. Here,
$C(q)=(2\sqrt{\pi})^{-1}\Gamma(q)/\Gamma(q-1/2)$, with $q=5/6$ being
the spectral index. For simplicity, the latter is assumed to be equal
for slab and 2D geometry. The remaining parameter in equations
(\ref{eq:pows}) and (\ref{eq:pow2d}) are the energy densities in slab
and 2D fluctuations, $\delta B_{slab}^2$ and $\delta B_{2D}^2$,
respectively, and the bend-over scales, denoted by $\lambda_{slab}$
and $\lambda_{2D}$. They are proportional to the respective
turbulence correlation lengths $l_{slab}$ and $l_{2D}$.

Concerning the correlation lengths, is assumed that
$l_{2D}=l_{slab}/10=10^9$m. Furthermore, it is used that the net
turbulent energy, $\delta B^2=\delta B_{slab}^2+\delta B_{2D}^2$, 
has only a small fraction in slab turbulent energy (20\%)
and is dominated by the 2D contribution (80\%), which is believed to
be consistent with solar wind observations (Bieber et al. \cite{bibetal94}). The
background magnetic field $B_0$ is given by $4\cdot 10^{-5}$ G, and
the ratio $\delta B/B_0$ is choosen to be $0.2$. For the decorrelation rate $\nu_c$, expression (\ref{eq:nudmt}) is employed (with
$\alpha_m=1$ and $v_A=50$ km s$^{-1}$) for both the slab and the 2D
contribution. However, in general, different parallel and perpendicular
decorrelation processes might govern particle diffusion. 

\begin{figure}
\includegraphics[width=9cm]{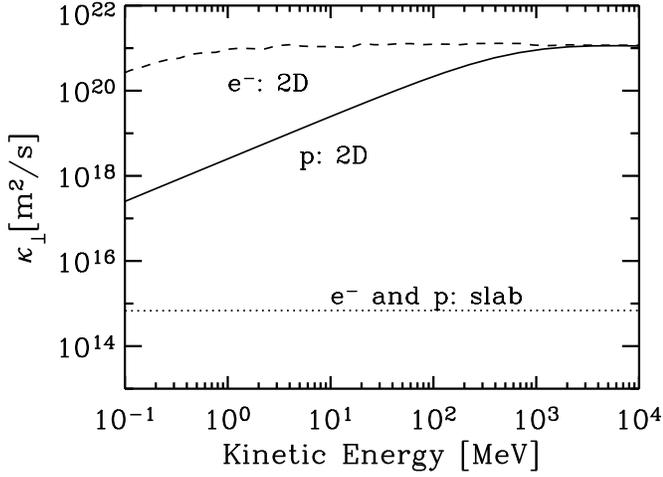}
\caption{Numerical solutions of equation (\ref{eq:kappadmt}) and
(\ref{eq:kappa2d}) representing the diffusion coefficients
  $\kappa_{\perp}^{slab}$ and $\kappa_{\perp}^{2D}$, respectively, as
  functions of the particle kinetic energy. The calculations were
  performed for protons (p) and electrons (e$^{-}$).}
\label{fig:fig2}
\end{figure}

Numerical results are shown in Figure \ref{fig:fig2}. The dotted line
gives the computations for $\kappa_{\perp}^{slab}$, equation
(\ref{eq:kappadmt}). Solutions for $\kappa_{\perp}^{2D}$,
Eq. (\ref{eq:kappa2d}), are visualized by the dashed and solid curves
for electrons (e$^-$) and protons (p), respectively. At a glance, for
the parameter used here,
quasilinear perpendicular diffusion is governed by the transverse
structure of the turbulent magnetic field. 

However, in view of recent numerical results by Giacaloni \&
Jokipii (\cite{giajok99}) (see their Fig. 7 for composite turbulence) and Matthaeus et al. (\cite{mat03}) for a
fully three-dimensional and static turbulence, one becomes aware of
the fact that QLT can not explain their simulations, in contrast to
the NLGC theory, where the transverse complexity of the turbulent
magnetic field plays a crucial role. For static conditions, the NLGC
theory provides for a nonvanishing, finite diffusion coefficient (see
Matthaeus et al. \cite{mat03}, Eq. [7] with $\gamma({\bf k})=0$). This is not the case in QLT,
as it is shown in section \ref{sec:slab} and \ref{sec:2d} for slab and
2D turbulence geometry, respectively, since $\kappa_{\perp}^{slab}$
vanishes and $\kappa_{\perp}^{2D}\to \infty$ for the magnetostatic case. This
implies for a fully three-dimensional turbulence
$\kappa_{\perp}=\kappa_{\perp}^{slab}+\kappa_{\perp}^{2D}\to\infty$.
QLT can, therefore, not provide for an adequate description for perpendicular
diffusion, at least for static turbulence. As shown above, QLT leads
to a nonvanishing, finite diffusion coefficient only if the turbulence
is non-static. In QLT, parallel and perpendicular diffusion processes
are considered as being independent, and the success of the NLGC model
is based on the nonlinear coupling of these two processes. It becomes
clear that such nonlinear effects are crucial for a more realistic
understanding of spatial particle diffusion and other processes such
as particle drift. 

Another issue concerns the transport of charged particles in
turbulent fields having at least one ignorable coordinate. It was
shown by Jokipii et al. (\cite{jok93}) and Jones et
al. (\cite{jonesetal1998apj}) that perpendicular particle diffusion is
then suppressed and that a particle is tied to the background magnetic
field. This result implies that it is not possible to get cross-field
diffusion for neither slab nor 2D turbulence geometry. The numerical
simulations by Qin et al. (\cite{qinetal2002grl}) were performed for a
``quasi''-slab, magnetostatic turbulence, leading to the insight that
perpendicular diffusion is indeed suppressed for such a simplified
geometry. The particle transport is then subdiffusive and not a
Markovian diffusion process. This can not be taken into account by
neither the NLGC model nor the QLT calculations presented here, since
both are based on the (Markovian) diffusion approximation. However,
though normal diffusion is explicitly assumed by using the TGK formula
(\ref{eq:tgk1}), the calculations in section \ref{sec:slab} show that
the QLT perpendicular diffusion coefficient approaches zero, for both
the plasma wave and the dynamical magnetic turbulence viewpoint, if
slab fluctuations are static. Giacalone \& Jokipii (\cite{giajok94})
have shown that cross-field diffusion is also suppressed for a static,
two-dimensional turbulence. This is in agreement with Jokipii et
al. (\cite{jok93}) and Jones et al. (\cite{jonesetal1998apj}). In
stark contrast to this is the QLT result for a two-dimensional
turbulence presented in section \ref{sec:2d}. There, it is shown that
$\kappa_{\perp}^{2D}\to\infty$ for static conditions.

The calculations presented here casts serious doubts on the
applicability of QLT for turbulence topologies different from slab
geometry. This implies that QLT is inapplicable for obtaining
conclusions on the nature of interplanetary magnetic fluctuations,
being in contradiction to the statement made by Shalchi \& Schlickeiser (\cite{shalchischlickeiser2004}).

Besides being valid only for weak turbulence ($\delta
B/B_0\ll 1$) and unperturbed particles orbits, the list of limitations
of QLT can be extended by the result that it fails for non-slab
geometries, at least for a static turbulence. Concerning the dynamical
behavior of the turbulence, Appendix \ref{sec:nlgc} describes briefly a
modification of the QLT results given above, yielding a
nonvanishing and finite diffusion coefficient for an ``intrinsic'' static
turbulence. The crude approach used there might probably be of more
academic interest, since it is based on the speculation that particles
alter the dynamical behavior of their scattering agent governing their
diffusive transport. However, it emphasizes the importance of
nonlinear contributions for perpendicular diffusion.

\begin{acknowledgements}
The author thanks R. A. Burger, J. Minnie and M. S. Potgieter for valuable conversations. Partial support by the South African National Research Foundation (NRF) and the Deutsche Forschungsgemeinschaft (DFG), SFB 591, is acknowledged.
\end{acknowledgements}

\appendix
\section{Derivation of $\kappa_{\perp}^{2D}$}
\label{sec:app}
To derive the quasilinear diffusion coefficient $\kappa_{\perp}^{2D}$, equation
(\ref{eq:kappa2d}), for 2D turbulence geometry, one can rearrange some
terms in the corresponding Fokker-Planck coefficient
(\ref{eq:nonres}), yielding
\begin{equation}
D_{\perp}^{2D}=\frac{2\pi R_L^2}{B_0^2}
\int\limits_{0}^{\infty}dk_{\perp}g(k_{\perp})\nu_c
\left[\frac{(1-\mu^2)}{2}\sum\limits_{n=-\infty}^{\infty}\frac{J_{n+1}^2(W)+J_{n-1}^2(W)}{z^2+n^2}+\mu^2\sum\limits_{n=-\infty}^{\infty}\frac{J_n^2(W)}{z^2+n^2}\right]
\label{eq:ap1}
\end{equation}
where $z=\nu_c/\Omega$. This corresponds to Eq. (54) of Shalchi \&
Schlickeiser (\cite{shalchischlickeiser2004}), but note that equation
(\ref{eq:ap1}) is derived from the general coefficient
(\ref{eq:dperp}). The two terms including the $n$-sum can
be subjected to a further treatment. Following Shalchi \&
Schlickeiser (\cite{shalchischlickeiser2004}), the formulas  
\begin{equation}
J_{n-1}^2(W)=\frac{2}{\pi}(-1)^{n-1}\int\limits_{0}^{\pi/2}d\theta J_0(2W\cos\theta)\cos(2\theta [n-1])
\end{equation}
and
\begin{equation}
J_{n+1}^2(W)=\frac{2}{\pi}(-1)^{n+1}\int\limits_{0}^{\pi/2}d\theta J_0(2W\cos\theta)\cos(2\theta [n+1])
\end{equation}
can be used (see Gradshteyn \& Ryzhik \cite{gradryz1966}), so that
\begin{equation}
J_{n+1}^2(W)+J_{n-1}^2(W)=\frac{4}{\pi}(-1)^{(n+1)}\int\limits_{0}^{\pi/2}d\theta
J_0(2W\cos\theta)\cos(2\theta n)\cos(2\theta)
\end{equation}
The first term then reads
\begin{equation}
\frac{1}{2}\sum\limits_{n=-\infty}^{\infty}\frac{J_{n+1}^2(W)+J_{n-1}^2(W)}{z^2+n^2}=-\frac{2}{z\sinh(\pi
  z)}\int\limits_{0}^{\pi/2}d\theta
  J_0(2W\cos\theta)\cos(2\theta)\cosh(2\theta z)
\label{eq:ap2}
\end{equation}
where the following relation (Gradshteyn \& Ryzhik \cite{gradryz1966})
is used:
\begin{equation}
\sum\limits_{n=1}^{\infty}\frac{(-1)^n}{z^2+n^2}\cos(2\theta
n)=\frac{\pi}{2z}\frac{\cosh(2\theta z)}{\sinh(\pi z)}-\frac{1}{2z^2}
\end{equation}
The second term can be treated in a similar manner, resulting in
\begin{equation}
\sum\limits_{n=-\infty}^{\infty}\frac{J_n^2(W)}{z^2+n^2}=\frac{2}{z\sinh(\pi
  z)}\int\limits_{0}^{\pi/2}d\theta
  J_0(2W\cos\theta)\cosh(2\theta z)
\label{eq:ap3}
\end{equation}
In order to solve the remaining $\theta$-integrations approximately, Shalchi \&
Schlickeiser (\cite{shalchischlickeiser2004}) used simplified
expressions for the power spectrum and the turbulence correlation
timescale. This is not done here. Instead, the integration with
respect to $\mu$ is first performed. To do so, equations
(\ref{eq:ap2}) and (\ref{eq:ap3}) are first inserted into 
(\ref{eq:ap1}). Making use of equation (\ref{eq:kappaperp}), one obtains
\begin{equation}
\kappa_{\perp}^{2D}=\frac{4\pi vR_L}{B_0^2}
\int\limits_{0}^{\infty}dk_{\perp}\frac{g(k_{\perp})}{\sinh(\pi z)}
\int\limits_{0}^{\pi/2}d\theta\cosh(2\theta z)I_{\mu}(\theta,\zeta)
\label{eq:ap4}
\end{equation}
with 
\begin{equation}
I_{\mu}(\theta,\zeta)=\int\limits_{0}^{1}d\mu J_0(2\zeta\cos\theta\sqrt{1-\mu^2})[1-2(1-\mu^2)\cos^2\theta]
\end{equation}
where $W=k_{\perp}R_L\sqrt{1-\mu^2}=\zeta\sqrt{1-\mu^2}$ is used. The
$\mu$-integration can be solved analytically (Gradshteyn \& Ryzhik
\cite{gradryz1966}) to obtain
\begin{eqnarray}
I_{\mu}(\theta,\zeta) &=&
(1-2\cos^2\theta)\int\limits_{0}^{\pi/2}d\phi\sin\phi J_0(2\zeta\cos\theta\sin\phi)+2\cos^2\theta\int\limits_{0}^{\pi/2}d\phi\sin\phi\cos^2\theta J_0(2\zeta\cos\theta\sin\phi)
\nonumber
\\
&=&
(1-2\cos^2\theta)\frac{\Gamma(1/2)}{\sqrt{4\zeta\cos\theta}}J_{\frac{1}{2}}(2\zeta\cos\theta)+\cos^2\theta\frac{\Gamma(3/2)}{(\zeta\cos\theta)^{3/2}}J_{\frac{3}{2}}(2\zeta\cos\theta)
\nonumber
\\
&=&
\frac{1}{4\zeta^3\cos\theta}\biggl[(1-2\zeta^2\cos(2\theta))\sin(2\zeta\cos\theta)-2\zeta\cos\theta\cos(2\zeta\cos\theta)\biggr]
\end{eqnarray}
where $\Gamma(x)$ and $J_{\frac{1}{2}+n}(x)$ denote the Gamma function
and spherical Bessel functions of the first kind, respectively.
With this, the diffusion coefficient (\ref{eq:ap4}) can be written as
\begin{equation}
\kappa_{\perp}^{2D}=\frac{\pi vR_L}{B_0^2}\int\limits_{0}^{\infty}dk_{\perp}g(k_{\perp})I(\zeta,z)
\label{eq:ap5}
\end{equation}
with the function $I(\zeta,z)$ given by equation (\ref{eq:funcI}).

\section{Modification of QLT Calculations}
\label{sec:nlgc}
In view of the failure of QLT in explaining perpendicular diffusion
for a static turbulence, the question arises whether the
quasilinear calculations presented above can be modified to obtain a
nonvanishing, finite diffusion coefficient. For this, the general
perpendicular diffusion coefficient is rewritten, and the DMT
viewpoint (damping model) is used. Upon substituting the general
Fokker-Planck coefficient (\ref{eq:dperp}) into $\kappa_{\perp}$,
Eq. (\ref{eq:kappaperp}), and using the DMT resonance function
(\ref{eq:dmtres}), one obtains
\begin{equation}
\kappa_{\perp}=\frac{v^2}{2B_0^2}
\sum\limits_{n=-\infty}^{\infty}
\int\limits_{-1}^{+1}d\mu
\int\limits_{}^{}d^3 k \frac{\nu_c}{\nu_c^2+(k_{\|}v_{\|}+n\Omega)^2}U({\bf k},\mu,n)
\label{eq:kappadamp}
\end{equation}
where $\nu_c=\tau_c^{-1}$ denotes again the turbulence decorrelation rate. 
Furthermore,
\begin{equation}
U({\bf k},\mu,n)=\frac{A({\bf k})}{k^2}\left[
\left((k_{\|}v\mu+n\Omega)^2v^{-2}+\mu^2k^2\right)J_{n}^2(W)+(1-\mu^2)k_{\perp}^2\bigl[J_{n}^{\prime}(W)\bigr]^2\right]
\label{eq:ufunc}
\end{equation}
Here, Eqs. (\ref{eq:fperph}), (\ref{eq:gperph}) and (\ref{eq:hperph})
were evaluated for $\omega_{j,R}=0$ and $\Gamma_j=0$ (DMT approach)
and inserted into equation (\ref{eq:dperp}). It was
assumed that $\sigma=0$.

Speculated that a number of possible effects may contribute to the 
decorrelation  of magnetic fluctuations. Particles diffuse in parallel and
perpendicular direction and interact with the turbulence and,
therefore, have probably an influence on the dynamics of their
scattering agent governing their diffusive transport. Proceeding, it is assumed that the
additional ``diffusive feedback'' of the particles on the turbulence
is given by $\kappa_{\|}$, the parallel diffusion coefficient, and by
$\kappa_{\perp}$ itself. This leads for the net turbulence decorrelation
rate to the Ansatz
\begin{equation}
\nu_c=\sum\limits_{i}\nu_{c,i}=\gamma+\kappa_{\|}k_{\|}^2+\kappa_{\perp}k_{\perp}^2+\ldots
\label{eq:assumption}
\end{equation}
where $\gamma({\bf k})$ now represents the intrinsic  turbulence
decorrelation rate. The dots denote other processes such as particle
drift and stochastic acceleration which might also influence turbulent
decorrelation and, therefore, perpendicular diffusion. In what
follows, equation (\ref{eq:kappadamp}) can be written as
\begin{equation}
\kappa_{\perp}=\frac{v^2}{2B_0^2}\sum\limits_{n=-\infty}^{\infty}
\int\limits_{-1}^{+1}d\mu
\int\limits_{}^{}d^3 k\frac{U({\bf k},\mu,n)}{(v/\lambda_{\|})\Lambda+\kappa_{\|}k_{\|}^2+\kappa_{\perp}k_{\perp}^2+\gamma}
\label{eq:kappanon}
\end{equation}
where the following auxiliary function is introduced:
\begin{equation}
\Lambda(\mu,n)=\frac{3(\mu+n/(k_{\|}R_L))^2}{1+(\kappa_{\perp}/\kappa_{\|})(k_{\perp}/k_{\|})^2+\gamma/\kappa_{\|}k_{\|}^2}
\end{equation}
It contains the QLT limit of an unperturbed particle orbit. Here, the
relation $\kappa_{\|}=\lambda_{\|}v/3$ has been used, with 
$\lambda_{\|}$ being the mean free path for parallel scattering. 
Apart from the terms appearing in $\Lambda$, equation
(\ref{eq:kappanon}) reveals the same nonlinear structure as the recent
NLGC result by Matthaeus et al. (\cite{mat03}) (see their Eq. [7]).

Going to the last extrem, it is assumed that particles move only
forward and backward to the mean magnetic field. This can be taken
into account in equation (\ref{eq:kappanon}) by considering the cases 
$\mu=\pm 1$ in terms of a Dirac delta distribution. Since $W=0$ for
$\mu=\pm 1$, the Bessel functions in Eq. (\ref{eq:ufunc}) are
nonvanishing only for $n=0$. The evaluation then yields
\begin{eqnarray}
\kappa_{\perp} & = & \frac{v^2}{B_0^2}
\int\limits_{}^{}d^3 k\frac{A({\bf
	k})(1+2k_{\|}^2/k_{\perp}^2)/(1+k_{\|}^2/k_{\perp}^2)}{(v/\lambda_{\|})\Lambda+\kappa_{\|}k_{\|}^2+\kappa_{\perp}k_{\perp}^2+\gamma}
\nonumber
\\
& \approx & \frac{v^2}{B_0^2}
\int\limits_{}^{}d^3 k\frac{A({\bf k})}{(v/\lambda_{\|})+\kappa_{\|}k_{\|}^2+\kappa_{\perp}k_{\perp}^2+\gamma}
\label{eq:kappanon2}
\end{eqnarray}
where, for the last step, the denominator is subjected to a rough
approximation and $\Lambda(\mu=\pm 1,n=0):=1$ is used. A closer
inspection of Eq. (\ref{eq:kappanon2}) leads to the result that the diffusion
coefficient is nonvanishing and finite for $\gamma=0$. Apart from
constants, Eq. (\ref{eq:kappanon2}) corresponds to the NLGC result by
Matthaeus et al. (\cite{mat03}). However, one should keep in mind
that the derivation of the diffusion coefficient (\ref{eq:kappanon2})
is based on a speculative approach and rough assumptions, and that it
is presented here for illustrative purposes only.


\begin{thebibliography}{}

\bibitem[1997]{biemat97}
Bieber, J. W., \& Matthaeus, W. H., 1997, \apj, 485, 655

\bibitem[1994]{bibetal94}
Bieber, J. W., Matthaeus, W. H., Smith, C. W. et al. 1994, \apj, 420, 294


\bibitem[1974]{for1}
Forman, M. A., Jokipii, J. R., \& Owens, A. J. 1974, \apj, 192, 535

\bibitem[1977]{for2}
Forman, M. A., 1977, \apss, 1, 83

\bibitem[1993]{gar93}
Gary, S. P., 2003, Cambridge Univ. Press, New York

\bibitem[1994]{giajok94}
Giacalone, J., \& Jokipii, J. R., 1994, \apj, 430, L137 

\bibitem[1999]{giajok99}
Giacalone, J., \& Jokipii, J. R., 1999, \apj, 520, 204

\bibitem[1969]{gle}
Gleeson, L. J., 1969, \planss, 17, 31


\bibitem[1966]{gradryz1966}
Gradshteyn, I.S., \& Ryzhik, I.M., 1966, Academic Press, New York


\bibitem[1966]{jo1}
Jokipii, J. R. 1966, \apj, 146, 480

\bibitem[1969]{jo2}
Jokipii, J. R., \& Parker, E. N., 1969, \apj, 160, 735

\bibitem[1993]{jok93}
Jokipii, J. R., K\'ota, J., \& Giacalone, J., 1993, \grl, 20, 1759

\bibitem[1990]{jones1990}
Jones, F.C., 1990, \apj, 361, 162

\bibitem[1998]{jonesetal1998apj}
Jones, F.C., Jokipii, J. R., \& Baring, M.G., 1998, \apj, 509, 238


\bibitem[2000]{kotajokipii2000}
K\'ota, J., \& Jokipii, J.R., 2000, \apj, 531, 1067


\bibitem[1957]{kubo1957}
Kubo, R., 1957, J. Phys. Soc. Japan, 12, 570

\bibitem[2001]{lersch01}
Lerche, I., \& Schlickeiser, R. 2001, \aap, 378, 279

\bibitem[2000]{mace}
Mace, R. L., Matthaeus, W. H., \& Bieber,J. W., 2000, \apj, 538, 192

\bibitem[1990]{mat90}
Matthaeus, W. H., Goldstein, M. L., \& Roberts, D. A., 1990, \jgr, 95, 20673

\bibitem[2003]{mat03}
Matthaeus, W. H., Qin, G., Bieber, J. W., \& Zank, G. P., 2003, \apj, 590, L53



\bibitem[2002a]{qin}
Qin, G., Matthaeus, W. H., \& and Bieber, J. W. 2002a, \apj, 578, L117

\bibitem[2002b]{qinetal2002grl}
Qin, G., Matthaeus, W. H., \& and Bieber, J.W., 2002b, \grl, 29, 1048

\bibitem[2002]{sch02}
Schlickeiser, R. 2002, Springer-Verlag, Berlin

\bibitem[2004]{shalchischlickeiser2004}
Shalchi, A., \& Schlickeiser, R., 2004, \aap, 420, 821

\bibitem[2001]{sta2}
Stawicki, O., Gary, S. P., \& Li, H., 2001, \jgr, 106, 8273

\bibitem[1977]{urch77}
Urch, I.H., 1977, AA\&SS, 46, 389
\end{thebibliography}
\end{document}